\documentstyle[pra,epsfig,floats,aps]{revtex}
\begin{document}

\title{Diffusion, localization and dispersion relations
 on ``small-world'' lattices}
\author{R{\'e}mi Monasson \cite{rm} }
\address{CNRS - Laboratoire de Physique Th{\'e}orique de l'ENS,
24 rue Lhomond, 75231 Paris cedex 05, France}
\date{\today}
\maketitle

\begin{abstract}
The spectral properties of the Laplacian operator on ``small-world''
lattices, that is mixtures of unidimensional chains and random graphs
structures are investigated numerically and analytically.  A transfer
matrix formalism including a self-consistent potential {\em \`a la}
Edwards is introduced. In the extended region of the spectrum, an
effective medium calculation provides the density of states and
pseudo relations of dispersion for the eigenmodes in close
agreement with the simulations. Localization effects, which are due
to connectivity fluctuations of the sites are shown to be
quantitatively described by the single defect approximation recently
introduced for random graphs.
\end{abstract}
\vskip .5 cm
\hskip 1cm PACS Numbers~: 63.50.+x-71.23.An-71.55.Jv
\vskip .5 cm

%
%

\section{Introduction}

Diffusion in random media is an an issue of great practical and
theoretical importance which has a received a lot of attention in the
past years \cite{ishi,sou,image,fractal,jp}. The presence of disorder,
e.g. impurities or random site energies, lattice defects, ... may
induce the localization of eigenstates and therefore have dramatic
consequences on transport properties \cite{sou,gen,and,thou}.  The
understanding of the localization transition, based on scaling theory
for finite-dimensional systems \cite{sca} has also greatly benefited
from the study of models on tree-like architecture \cite{abou}, that
is without any underlying geometry. Diffusion on such amorphous
structures has also attracted attention as a toy-model of dynamical
evolution in the phase space of complex, e.g. glassy systems
\cite{cam,bray}.

Recently, it was pointed out that mixed lattices, built from a mixture
of finite-dimensional structures and random graphs could be of
interest to modelize various real situations taking place in different
scientific fields \cite{stro}.  As a main attractive feature, these
``small-world'' models simultaneously combine short- and long-range
interactions and could display some new and interesting 
properties stemming from this unusual coexistence
\cite{stro,bart,alain}.

From a physical point of view, this situation is already encountered
in the field of polymers or more generally long molecules
\cite{deg}. Couplings between successive monomers give rise to a
unidimensional system while (e.g. hydrodynamics) interactions between
constituents in the three-dimensional space are of long range nature
along the polymeric chain. Of course, the long-range links formed this
way are strongly correlated and cannot be considered to be independent
as in the small-world paradigm.  Some successfull approximation
schemes developed in the context of polymer theory, e.g. the
self-consistent potential of Edwards \cite{edw,deg2} method precisely
amount to neglect such correlations and nevertheless obtain reliable
results \cite{deg}. Following Zimm's work \cite{zimm}, many studies
have concentrated on the influence of such interactions on the
relaxation dynamics of polymers \cite{deg,har}. However, these
calculations can in a sense be considered as {\em annealed} since the
hydrodynamical interactions tensor is averaged over chain positions
{\em before} normal modes are computed and not {\em later} as it
should. Such a pre-averaging approximation \cite{har} may lead to
qualitatively erroneous predictions especially for large eigenvalues 
and must be considered with caution.
Diffusion on the small-world lattice can therefore be viewed as an
elementary step toward the understanding of small times vibrations or
equivalently the instantaneous normal modes \cite{inm,stra,aut} of a
polymeric chain around a {\em quenched} configuration.

From a theoretical point of view, the small-world structure is also
worth being investigated since it mixes two structures which can be
studied in great mathematical details \cite{stro}. Diffusion on a
one-dimensional lattice obsviously reduces to the study of dispersion
relations for plane waves \cite{image}. As for the random graph
structure, a large attention has also been paid to its spectral
properties \cite{bray,ded,fie,sda,dav,fig}. The main physical feature is
the emergence of localized states centered on the geometrical defects
of the graph, that is sites with abnormally low or large
connectivities with respect to the average coordination number
\cite{sda}.  As we shall see, the combination of both structures can
be precisely studied and allows to answer some interesting questions
concerning localization and the existence of pseudo-dispersion
relations for the eigenmodes in presence of disorder \cite{gen,thor}.

This paper is organized as follows. In Section II, the small-world
model is defined and related to its building bricks, that is a random
graph superposed to a one-dimensional lattice.  The main features of
diffusion on the latters are briefly recalled and used to guess some
spectral properties of the small-world Laplacian. Section III exposes
the numerical results obtained from exact diagonalizations. The
density of states as well as some characteritics of the eigenmodes,
e.g. localization properties and autocorrelation functions are
analyzed. We present in Section IV the analytical approach to attack
the problem, based on a self-consistent transfer matrix formalism
\cite{edw,deg2}. The region of the spectrum corresponding to extended
eigenstates is studied in Section V by means of an effective medium
approximation \cite{ishi,thor}.  Localization properties are
unravelled in Section VI using a more refined scheme of approximation
exploiting the geometrical defect mechanism exposed above \cite{sda}.

%
%

\section{The small-world lattice and its structure}

\subsection{Presentation of the model}

We consider $N$ points $A_i$, $i=1,\ldots ,N$ on a one-dimensional
ring ${\cal R}$.  Each point $A_i$ is connected to its $2 K$ nearest
neighbors $A_{i\pm 1}, A_{i\pm 2}, \ldots , A_{i\pm K}$ (where site
labels are defined modulo $N$ to ensure periodicity on ${\cal R}$). In
addition, we superpose to ${\cal R}$ a random graph ${\cal G}$ with
mean connectivity $q$ defined as follows. Each pair of points $(A_i ,
A_j)$ (such that the distance $|i-j|$ on the ring is larger than $K$)
is connected with probability $q/N$ and left unchanged with
probability $1-q/N$.

The small-world lattice ${\cal S}$ is the union of both graphs ${\cal
R}$ and ${\cal G}$. In average, ${\cal S}$ includes $(K+q/2)N$ edges
to largest order in $N$. The coordination degree $c_i$ of $A_i$ is a
random variable bounded from below by $2K$. More precisely, $c_i - 2K$
obeys a Poisson law of parameter $q$. This ``small-world'' lattice
differs from the definition of \cite{stro} but maintains the
coexistence between short- and long-range links. In addition it leads
to simpler analytical calculations than the model exposed in
\cite{stro}.

We then consider the Laplacian operator $W^{\cal S}_{ij}$ on ${\cal
S}$. For $i\ne j$, $W^{\cal S}_{ij}=-1/2K$ if $A_i$ and $A_j$ are
connected, 0 otherwise. Diagonal elements, $W^{\cal S}_{ii}=c_i/2K$
ensure the conservation of probability. Note that, with respect to the
usual definition, eigenvalues are rescaled by a multiplicative factor
$-1/2K$ so that the support of the spectrum becomes positive.

\subsection{Definition of spectral quantities}

We call $\lambda _e$ (respectively $w _{j,e}$) the eigenvalues
(respectively the components of the associated eigenvectors normalized
to unity) of the Laplacian $W ^{\cal S}$, with $e=1,\ldots ,N$.  
Most spectral properties of $W^{\cal S}$ can be obtained through the
calculation of the resolvent \cite{thou}
\begin{equation}
G^{\cal S} _{jk} (\lambda + i \epsilon) = \bigg( (\lambda + i\epsilon
) 1-W^{\cal S} \bigg) ^{-1} _{jk} =
\sum _{e=1} ^N \frac{ w _{j,e} w _{k,e} }{ \lambda - \lambda _e + i
\epsilon } \qquad .
\label{fieldd}
\end{equation}
The mean density of eigenvalues indeed reads 
\begin{equation}
p (\lambda )= - \frac{1}{N\pi} \lim _{\epsilon \to 0 ^+} \sum _{j=1}^N 
\hbox{\rm Im} \;\overline{ G^{\cal S} _{jj}(\lambda + i \epsilon ) } \qquad , 
\label{spectre}
\end{equation}
where the overbar denotes the average over disorder, that is the
random graph ${\cal G}$. 

Another quantity of interest is the autocorrelation function of
eigenvectors. The power spectrum $|\tilde w _ e(\theta ) |^2$ of
eigenmode $e$, {\em i.e.} the squared modulus of its Fourier transform
reads
\begin{equation}
|\tilde w _e (\theta ) |^2 = \sum _{j,k =1}^N w_{j,e} w_{k,e} \exp(
2i\pi \theta (j-k)) \quad .
\label{beso}
\end{equation}
We then define $|\tilde w (\theta ,\lambda ) |^2$ as the sum
of $|\tilde w _e (\theta ) |^2 $ over all $w _{i,e}$ lying in the
range $\lambda \le \lambda _e \le \lambda + d\lambda$, divided by the
number $Np (\lambda ) d\lambda$ of such eigenvectors. Once ensemble
average is carried out, this power spectrum is simply related to the
off-diagonal resolvent (\ref{fieldd}) through
\begin{equation}
\frac{1}{N} \sum _{j,k=1}^N \overline{G^{\cal S} _{jk}(\lambda + i \epsilon )}
\exp( 2i\pi \theta (j-k)) = \int _0 ^\infty d\mu \; p(\mu )\; \frac{
\overline { |\tilde w (\theta ,\mu ) |^2} }{\lambda-\mu + i\epsilon}
\qquad ,
\label{rel1}
\end{equation}
where we have assumed that the density of states $p$ is self-averaging
\cite{pastur}, as confirmed by numerics when the size of the sample
$N$ becomes large.

Localization properties of eigenmodes can be studied through the
knowledge of inverse participation ratios \cite{sou,thou}. The inverse
participation ratios of eigenmodes $e$, defined as $w_e ^4 = \sum _{i}
|w _i^{(e)} |^4 $ are then averaged over a small energy region around
$\lambda$ to give $w^4 (\lambda )$ following the procedure described
above. Though inverse participation ratios may fluctuate from state to
state at a given eigenvalue $\lambda$, a non vanishing value of the
ensemble averaged $\overline{w^4 (\lambda )}$ gives a good indication
of the emergence of localization \cite{sou}.
 
The spectral properties of ${\cal S}$ will of course reflect its mixed
structure. Before turning to numerics, it is therefore useful to 
briefly recall known results on the spectral properties of both ${\cal
R}$ and ${\cal G}$ separately. 
  
\subsection{Diffusion on the unidimensional ring}

On the ring, the Laplacian operator $W ^{\cal R}$ is diagonalized by 
plane waves $w_\phi$ whose components at site $j$ read
\begin{equation}
(w_\phi) _j = \frac 1{\sqrt N} \; \exp \left( 2i \pi \phi j \right) \qquad ,
\label{wave1}
\end{equation}
where the phase $\phi$ assumes multiple values of $1/N$, from $0$ up
to $1-1/N$. In the large $N$ limit, $\phi$ becomes a continuous
variable ranging from $0\le \phi <1$ and the corresponding eigenvalue
is given by
\begin{equation}
\lambda ^{\cal R}(\phi) = 1 - \frac 1K \sum _{\ell =1}^K \cos (2 \pi \; \ell \;
\phi ) \qquad .
\label{disperpur}
\end{equation}
Due to the reflection symmetry $\phi \to 1-\phi$, we can restrict to
$\phi$ belonging to the interval $[0;1/2]$. The dispersion relation
(\ref{disperpur}) defines the inverse multivalued function $\phi
^{\cal R} (\lambda )$ plotted fig~1a for $K=3$. The eigenvalue
$\lambda$ is stationary for four different wave numbers $\phi^{\cal R}
_0 =0$, $\phi^{\cal R} _1 \simeq 0.206$, $\phi ^{\cal R} _2 \simeq
0.354$ and $\phi ^{\cal R}_3 =1/2$ corresponding to $\lambda ^{\cal
R}_0 =0$, $\lambda ^{\cal R}_1 \simeq 1.439$, $\lambda ^{\cal R}_2
\simeq 0.981$ and $\lambda ^{\cal R}_3 =4/3$. The density of states $p
^{\cal R}(\lambda )$ is given by
\begin{equation}
p^{\cal R}( \lambda ) =\sum _{\phi / \phi
^{\cal R} (\lambda ) = \phi} \frac
 1{| \lambda ^{\cal R}(\phi)'|} \qquad ,
\label{spectrepur}
\end{equation}
and is shown fig~1b for $K=3$. Van Hove singularities are located at
$\lambda ^{\cal R}_0$, $\lambda ^{\cal R}_1$, $\lambda ^{\cal R}_2$
and $\lambda ^{\cal R}_3$.

\begin{figure}
\includegraphics[width=200pt,angle=-90]{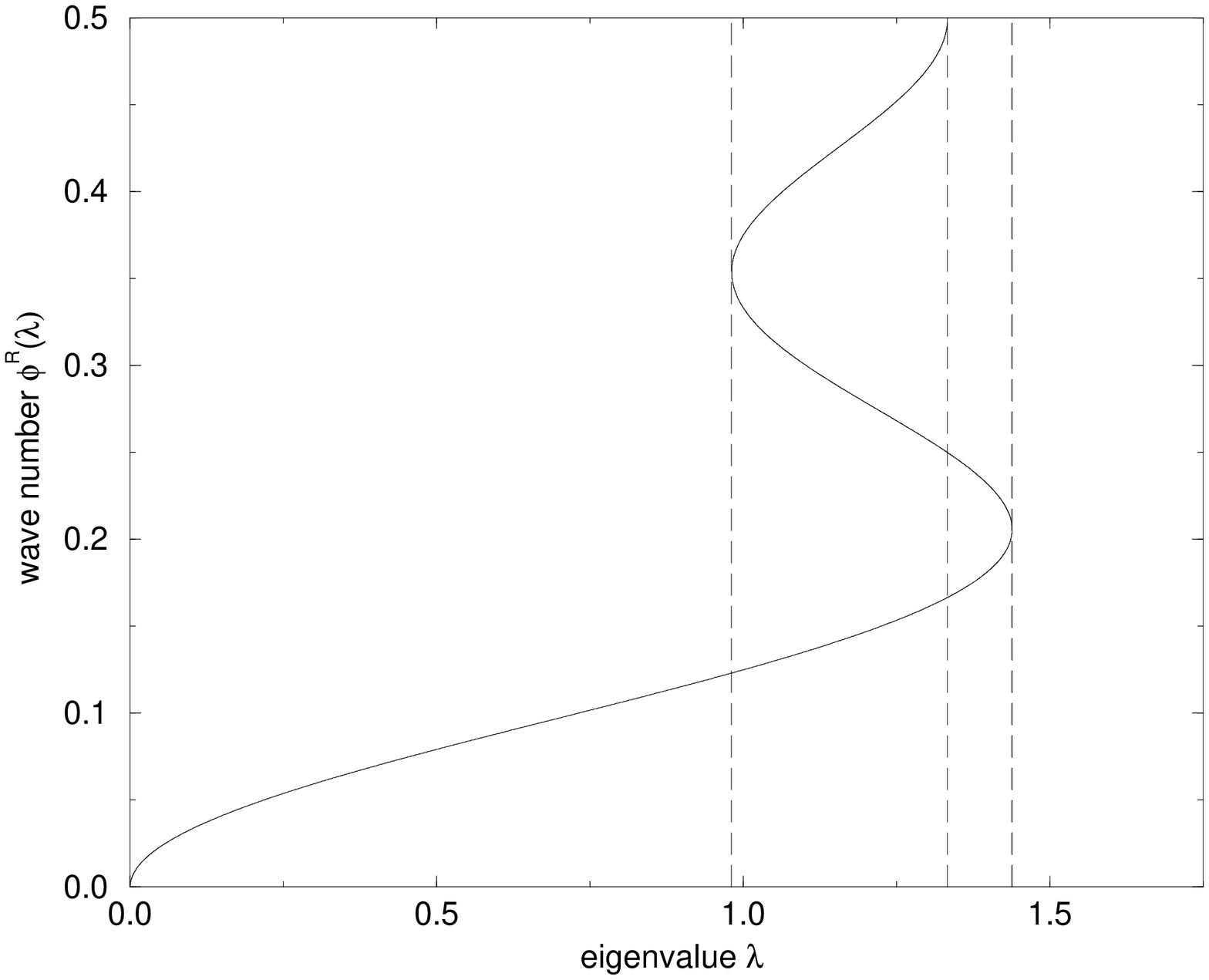}
\includegraphics[width=200pt,angle=-90]{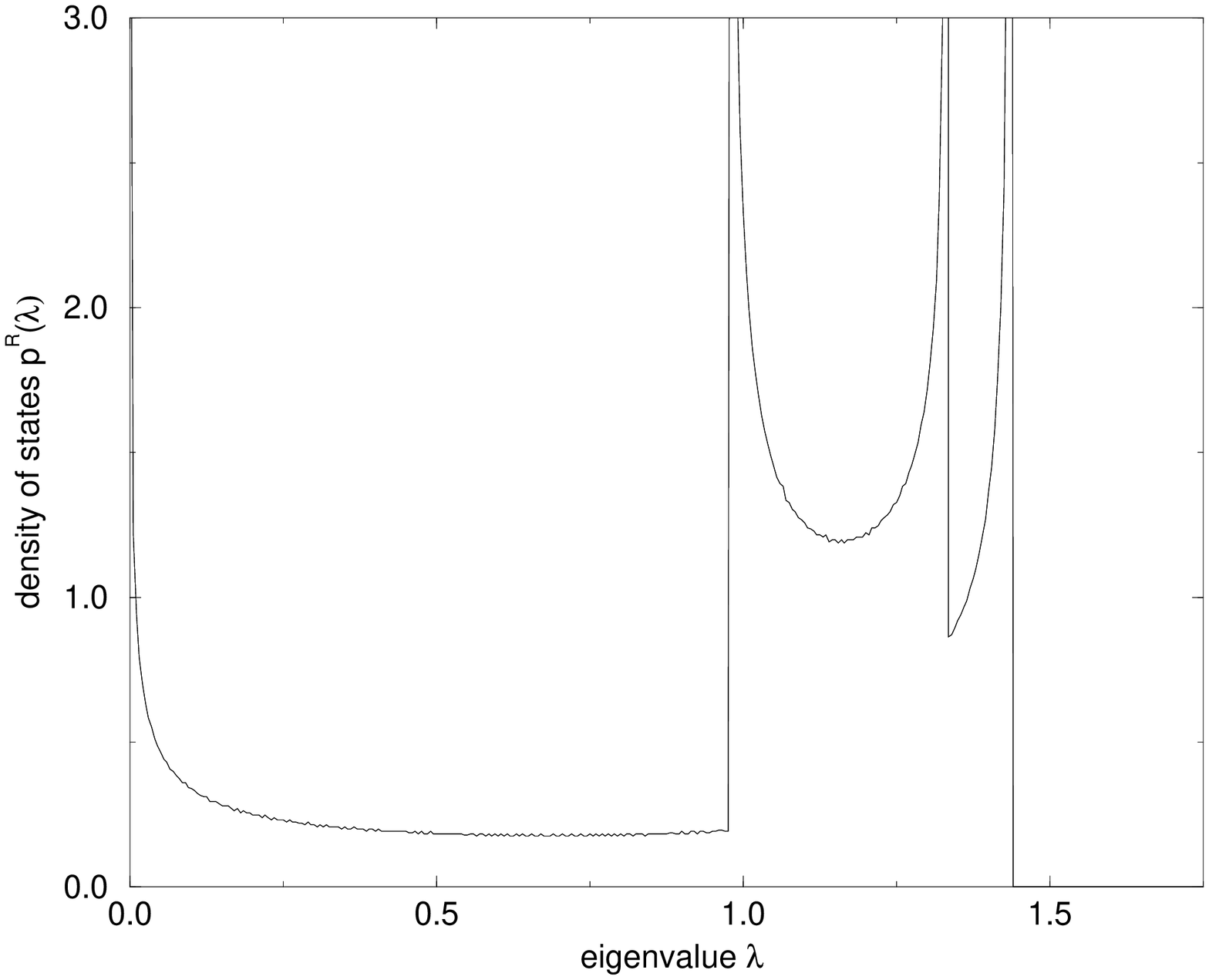}
\vskip 1cm
\caption{The $K=3$ ring. {\bf a}~: Relation of dispersion; the plane
wave phase $\phi (\lambda )$ is plotted as a function of the
eigenvalue $\lambda $(full curve). Non zero stationary eigenvalues are
indicated by dashed lines (from left to right~: $\lambda ^{\cal R}_2$,
$\lambda ^{\cal R}_3$ and $\lambda ^{\cal R} _1$, see text). {\bf b}~:
Density of states $p ^{\cal R}(\lambda )$ as a function of $\lambda $.
Van Hove singularities take place at $\lambda ^{\cal R}_0$, $\lambda
^{\cal R}_2$, $\lambda ^{\cal R}_3$ and $\lambda ^{\cal R}_1$ (from
left to right) } \protect\label{f1}
\end{figure}

The relation of dispersion may be found back by looking at the
autocorrelation function of eigenvectors, or alternatively at their
power spectrum $|\tilde w (\theta , \lambda ) |^2 $, see Section II.B.
According to fig~1a, $|\tilde w (\theta , \lambda ) |^2$ will exhibit
one (respectively two or three) Dirac peaks 
 over the range $0\le \theta\le \frac 12$ when $0<\lambda
<\lambda ^{\cal R}_2$ (respectively $\lambda ^{\cal R}_3<\lambda
<\lambda ^{\cal R}_1$ or $\lambda ^{\cal R}_2<\lambda <\lambda ^{\cal
R}_3$).  The locations of the peaks coincide with the multivalued
function $\phi ^{\cal R}(\lambda )$, giving back the relation of dispersion
(\ref{disperpur}). The study of the power spectrum will turn out to be
useful for establishing pseudo-dispersion relations on the small-world
lattice in presence of disorder.

\subsection{Diffusion on the random graph}

Diffusion on the random graph has been the object of several
analytical and/or numerical studies \cite{bray,ded,fie,sda,dav,fig}.
Briefly speaking, the two main physical features are~:
\begin{itemize}
\item The spectral density has a bell-like shape with small
oscillating tails (see fig~1 in \cite{sda}). The central bell-like
part corresponds to extended states.
\item The lateral oscillations are accompanied by increase of the
inverse participation ratio, which is an indication of the
localization of eigenmodes near mobility edges. Localized states are
centered on geometrical defects, that is sites with abnormally large
or low connectivities with respect to the average coordination number.
\end{itemize}
More quantitatively, defects will pin localized states if their
connectivity $c$ deviate from the average coordination number $q$ by
more than $\pm \sqrt q$.  Due to the defect mechanism, the localized
peak associated to connectivity $c$ has a total weight
(i.e. integrated density of states) well approximated by the Poisson
law $p^{c} e^{-p} / c!$.

\subsection{Perturbation theory and heuristic arguments}

We start by decomposing $W^{\cal S}$ into $W ^{\cal R} + W ^{\cal
G}$. One is tempted to treat $W^{\cal G}$ as a small perturbation with
respect to the ring Laplacian $W^{\cal R}$ when $q\ll 1$. The
resolvents $G^{\cal S}$ and $G^{\cal R}$ for the small-world lattice
and the ring obey the Dyson equation
\begin{equation}
G^{\cal S} _{jk} (\lambda +i \epsilon ) = 
G^{\cal R} _{jk} (\lambda +i \epsilon ) + \sum _{\ell ,m = 1} ^N
G^{\cal R} _{j\ell} (\lambda +i \epsilon ) \; W ^{\cal G} _{\ell m} \;
G^{\cal S} _{mk} (\lambda +i \epsilon ) 
\qquad .
\end{equation}
Using the above equation, the Fourier transform of the self-energy
$\Sigma$ at energy $\lambda$ and momentum $\phi$ can be computed order
by order within perturbation theory \cite{thou}.

At first order, the self-energy reads using (\ref{wave1}),
\begin{equation}
\Sigma _1 (\lambda + i\epsilon , \phi ) = \sum _{j,k=1}^N \overline{
W^{\cal G} _{jk} }\; (w_\phi) _j \; (w_\phi) ^* _k = \frac q{2K}
\qquad ,
\end{equation}
for non zero phases $\phi$. In other terms, the whole spectrum on
fig~1b would be shifted to the right by $q/2K$.

Second order corrections to the self-energy may be written as
\begin{equation}
\Sigma _2 (\lambda + i\epsilon , \phi ) = \sum _{j,k=1}^N \sum
_{\ell ,m=1}^N \overline{W^{\cal G} _{jk}  
W^{\cal G} _{\ell m} } ^{\; c} \; G ^{\cal R}_{k\ell }(\lambda +
i \epsilon ) \; (w_\phi) _j \; 
(w_\phi) ^* _m = \frac q{2K^2} \int _0
^\infty d\mu \; \frac{p^{\cal R}(\mu )}{\lambda -\mu + i\epsilon } 
\qquad ,
\end{equation}
where the subscript $c$ indicates the connected two-points correlation
function of $W^{\cal G}$.  The emergence of an imaginary part in the
self-energy shows that the resolvent $\overline{G^{\cal S}_{jk}}$ at
energy level $\lambda$ will decrease exponentially with the distance
$|j-k|$ over a typical length
\begin{equation}
L_E (\lambda ) = \frac{2 K^2}{\pi} \; q^{-1}  \; \big( p^{\cal R}
(\lambda ) \big) ^{-1}
\qquad . \label{le}
\end{equation}
The latter may be interpreted as the elastic mean path \cite{sou,sor}
travelled by a wave at energy level $\lambda$ between two successive
scattering events. In other words, the peaks appearing in the
power spectrum of the eigenvectors (\ref{beso}) will acquire some
finite width $\sim 1/L_E$, see Section IV.D. $L_E (\lambda )$
indeed diverges for weak disorder $q\to 0$. Clearly, perturbation
theory is sensible as long as the wave length $1/\phi ^{\cal R}
(\lambda ) $ keeps much smaller than the elastic mean path $L_E
(\lambda )$, the so-called Ioffe-Regel criterion \cite{sou}. This
condition obviously breaks down at van Hove singularities, see Section
II.C. Furthermore, at moderately large eigenvalues ($\lambda > \lambda
^{\cal R} _2$ in the $K=3$ case), the existence of several distinct
branches in the dispersion relation (fig~1a) makes the density of
states higher than at smaller eigenvalues ($\lambda < \lambda ^{\cal
R} _2$) as seen fig~1b. We therefore expect from 
expression (\ref{le}) that disorder will affect the
density of states at large eigenvalues more strongly than at small
energy levels.

This reasoning should hold except at vanishing eigenvalues
$\lambda$. One-dimensional model indeed present the peculiarity that
the density of states diverges at zero energy. More precisely, 
$L_E (\lambda ) \sim \sqrt \lambda $ as $\lambda \to 0$ and therefore
perturbation theory should not be trust below $\lambda = O(q)$. 

What happens at smaller energy levels? Some heuristic arguments were
developed in \cite{bray} to answer this equation in the random graph case. 
First of all, calculations of the spectrum on a Cayley tree with {\em fixed}
connectivity $c$ show a gap between the null eigenvalue and the left
side of the spectrum support which disappear in the limit $c\to 2$
only. Indeed, long enough chain-like structure are capable of exhibiting
small eigenvalues of the order of their inverse squared length
\cite{bray}. It is appealing but speculative to think that the
small eigenvalues of $W^{\cal S}$ come from rare portions of the 1-D ring
unaltered by disorder \cite{bray}. This Lifshitz-like argument may be 
quantified as follows. A chain of $M$ sites, giving rise to
eigenvalues of the order of $1/M^2$  will not be corrupted by
the random graph component ${\cal G}$ with an exponentially small
probability scaling as $\exp( -q M)$. Summing these contributions to
the density of states over large $M$ values, we obtain
\begin{equation}
p(\lambda ) \simeq  p^{\cal R} (\lambda ) \; \exp \left(- \frac 
{q}{\phi ^{\cal R} (\lambda ) } \right) \qquad ,
\label{br}
\end{equation}
for small $\lambda$'s. Expression (\ref{br}) has a nice interpretation
in terms of the disordered-induced length $L_d = 1/q$. The eigenmodes of
the ring are exponentially attenuated with the ratio
of their wave-length over the typical distance $L_d$ between two successive
random links incoming onto the one-dimensional structure. When
$\lambda \to 0$, the exponential term in (\ref{br}) vanishes as $\exp
( - C q / \sqrt \lambda )$ (where $C$ is a constant) and thus screens
the algebric divergence of $p^{\cal R}$. This argument predicts a
cross-over between the pure case and the Lifshitz tail at $\lambda
_{c.o.} = O(q^2)$ corresponding to a density $p_{c.o.} =
O(q^{-1})$. We shall see in next Sections that such a cross-over is
indeed present.

%
%

\section{Numerical diagonalization}

In this Section we restrict to $K=3$. The corresponding ring has a non
trivial relation of dispersion giving rise to richer features
than for smaller values of the connectivity, e.g. $K=1$. In addition,
diagonalizations of small-world lattices with larger $K$ suffer from bigger
finite-size effects and thus reliable results demand a 
prohibitive computing cost. 

\subsection{Description of the eigenvalues spectrum}

\begin{figure}
\includegraphics[width=200pt,angle=-90]{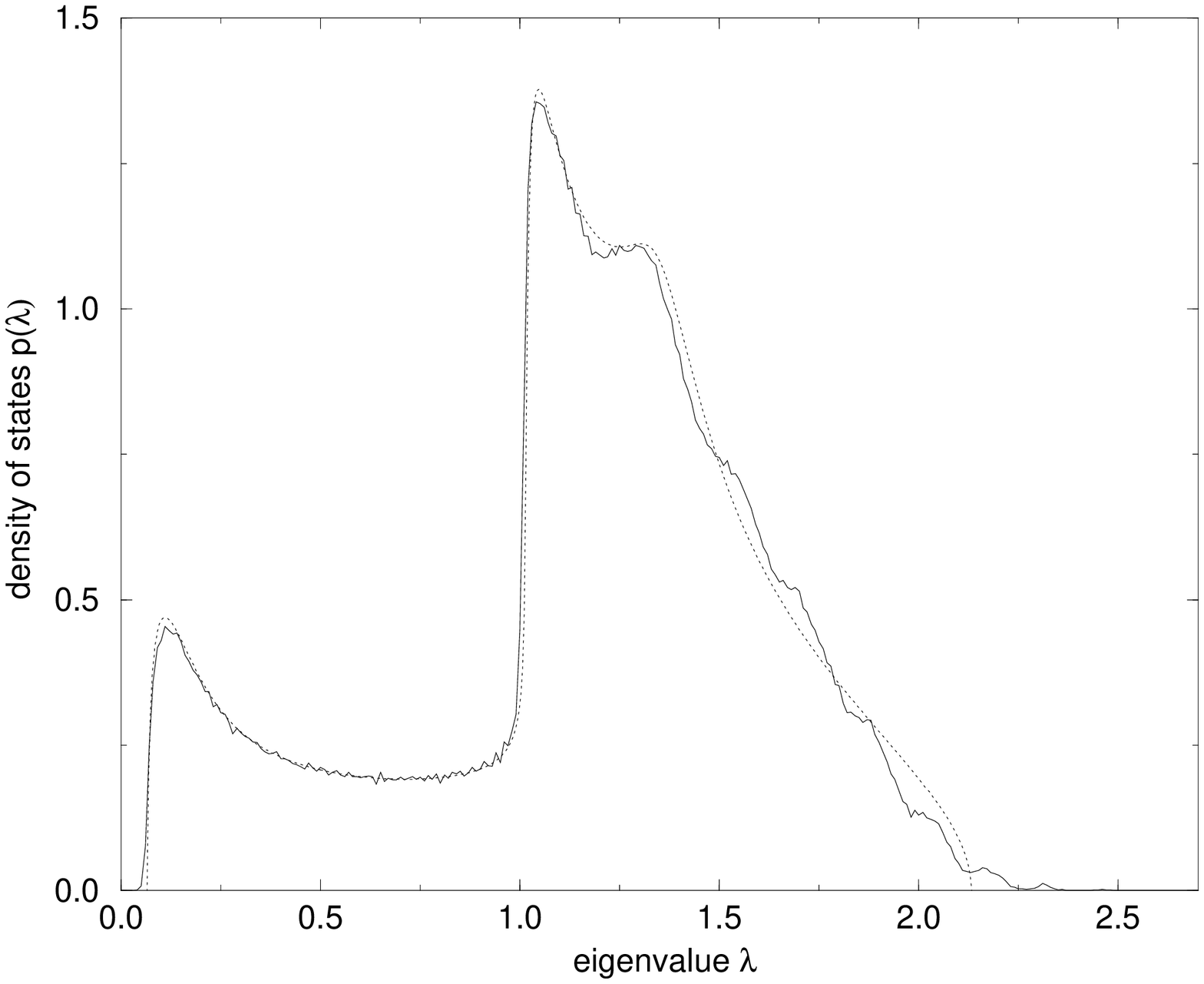}
\includegraphics[width=200pt,angle=-90]{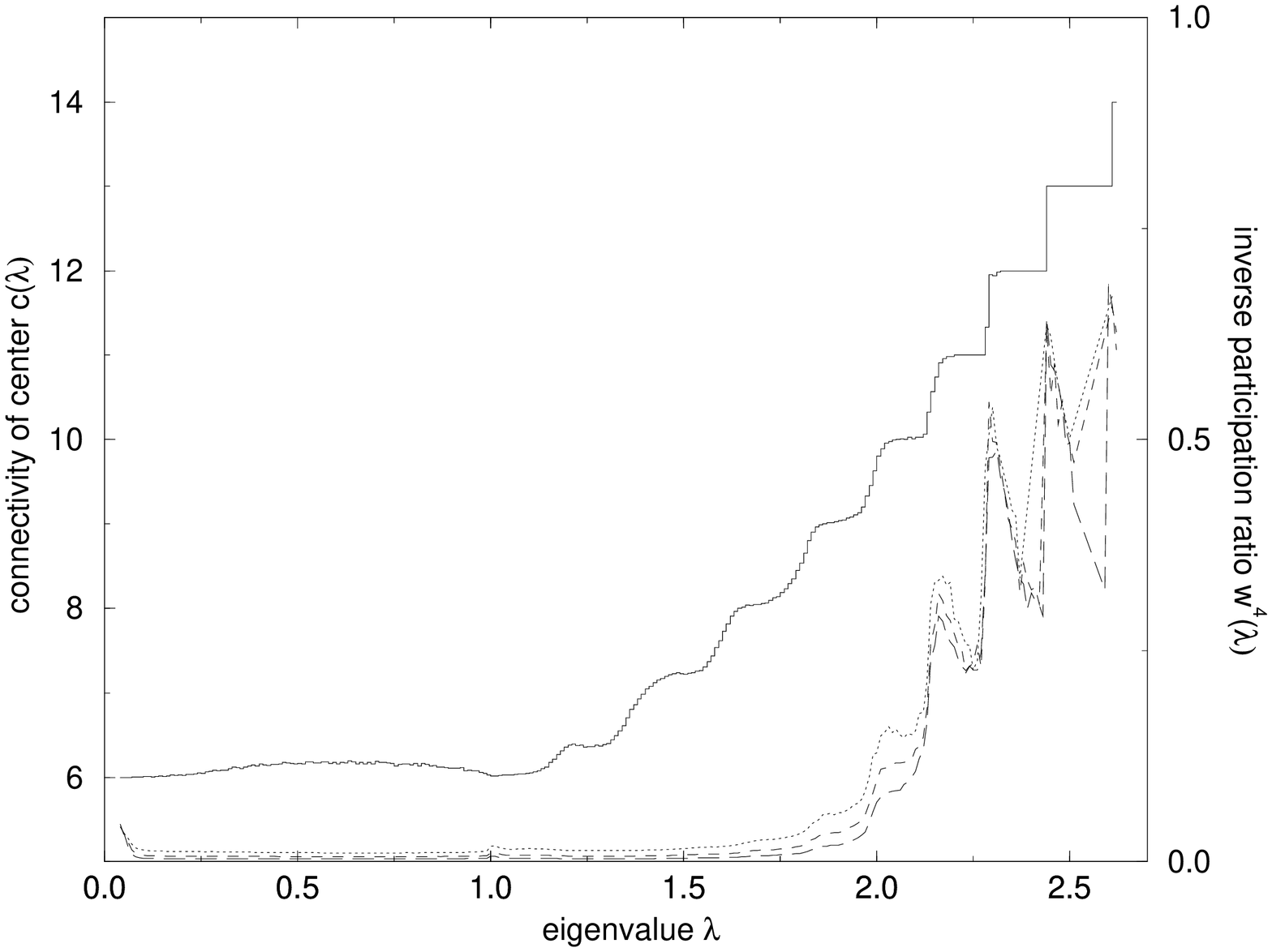}
\vskip 1cm
\caption{Results for connectivity parameters $K=3$, $q=1$.
{\bf a}~: 
density of states $p$ obtained from numerics (full curve) and EMA (dotted
line) approximation.
{\bf b}~: coordination of eigenstate-center $c$ (full curve) and 
inverse participation ratio $w^4$ for $N=256$ (dotted curve), $N=512$
(dashed curve) and $N=1000$ (dot-dashed curve) averaged over 1000
samples.   } 
\protect
\end{figure} 

We have performed exact numerical diagonalizations of the random
Laplacian $W^{\cal S}$ in the case $K=3$ for different sizes, up to $N=1000$
and for 1000 samples.  Fig~2a and
fig~2b respectively display $p (\lambda )$ and $w^4 (\lambda )$ for
connectivity parameters $K=3,q=1$.  Main remarks are in order~:
\begin{itemize}
\item the density of states at small eigenvalues seems to vanish
for $\lambda < \lambda_- \simeq 0.045\pm 0.005$. There is no true gap
strictly speaking but rather a very small density of states 
(discussed in Section II.E) that
cannot be accounted for by numerics due to the statistical shortfall
of eigenvalues. In the following, the $[0;\lambda_- ]$ range will be
referred to as a ``pseudo-gap''.
\item the density of states exhibit a maximum on the left side of the
spectrum at $\lambda _{c.o.} \simeq 0.11$, slightly above $\lambda
_-$. 
\item the central part of the spectrum ($\lambda _{c.o.} < \lambda <
\lambda_+ $) has a smooth shape and corresponds to extended
states. The general form of $p(\lambda )$ is reminiscent of the
density of states for the ring fig~1b but divergences have been
smeared out by disorder \cite{gen}.  Notice that in the range $\lambda
_{c.o.} <
\lambda < 1$, the density of states is in quantitative agreement with
the ring spectrum, compare fig~1b and fig~2a.  For increasing sizes
$N$ and at fixed $\lambda$, $w^4$ vanishes as $1/N$ and the breakdown
of this scaling identifies the upper mobility edge~: $\lambda _+
\simeq 2.1 \pm 0.1 $.
\item on the right side of the spectrum, the density exhibits
successive regular peaks and the eigenstates become
localized, see fig~2b. 
\end{itemize}

Similar results are shown fig~3a-b for $K=3$ and a larger random graph
component $q=5$. In this case, we have found $\lambda _-\simeq 0.4 \pm
0.05 $, $\lambda _{c.o.} \simeq 0.7 \pm
0.05 $ and $\lambda _+ \simeq 3.3 \pm 0.1 $. As expected, localization
effects are stronger and the pseudo-gap becomes wider.

\begin{figure}
\includegraphics[width=200pt,angle=-90]{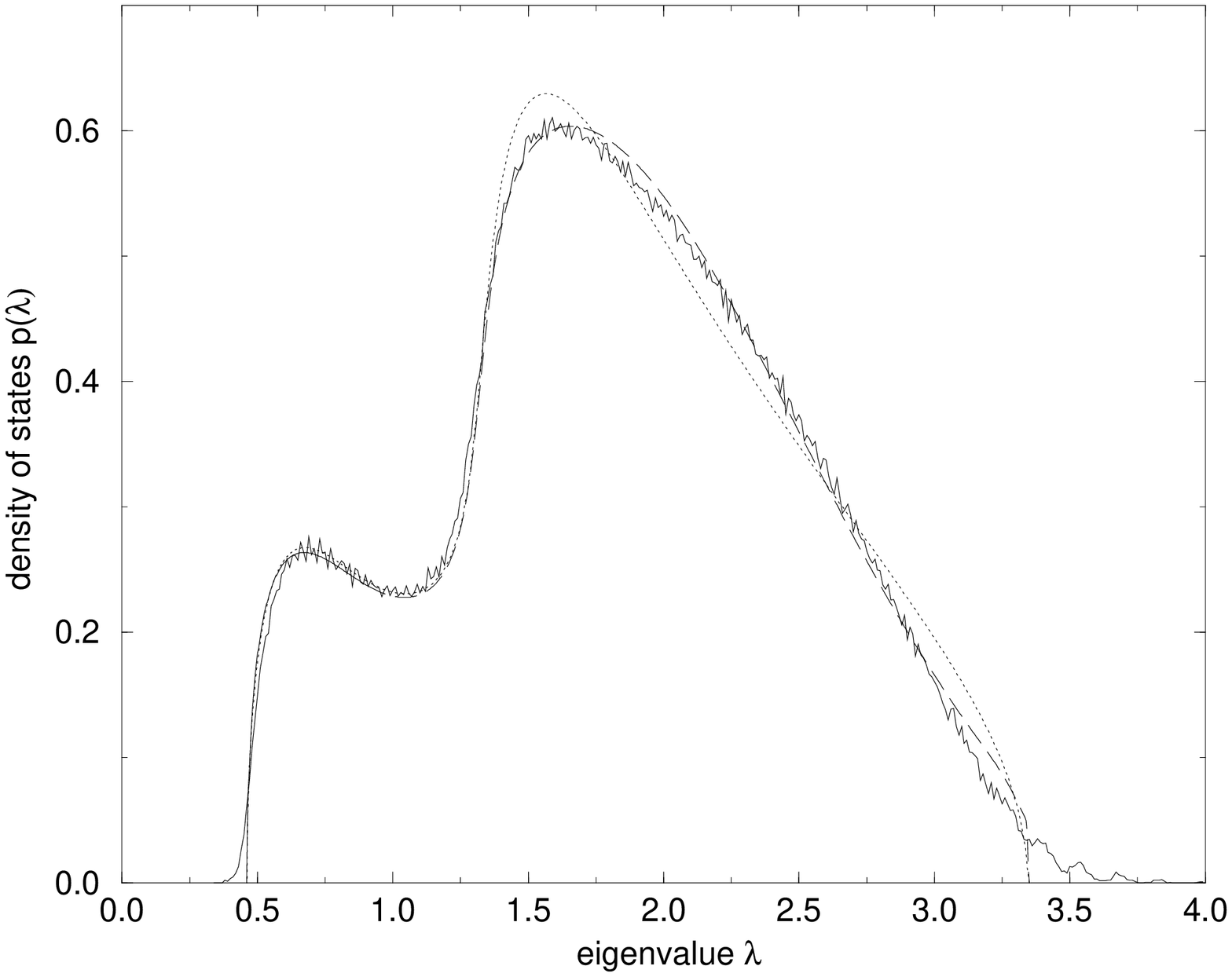}
\includegraphics[width=200pt,angle=-90]{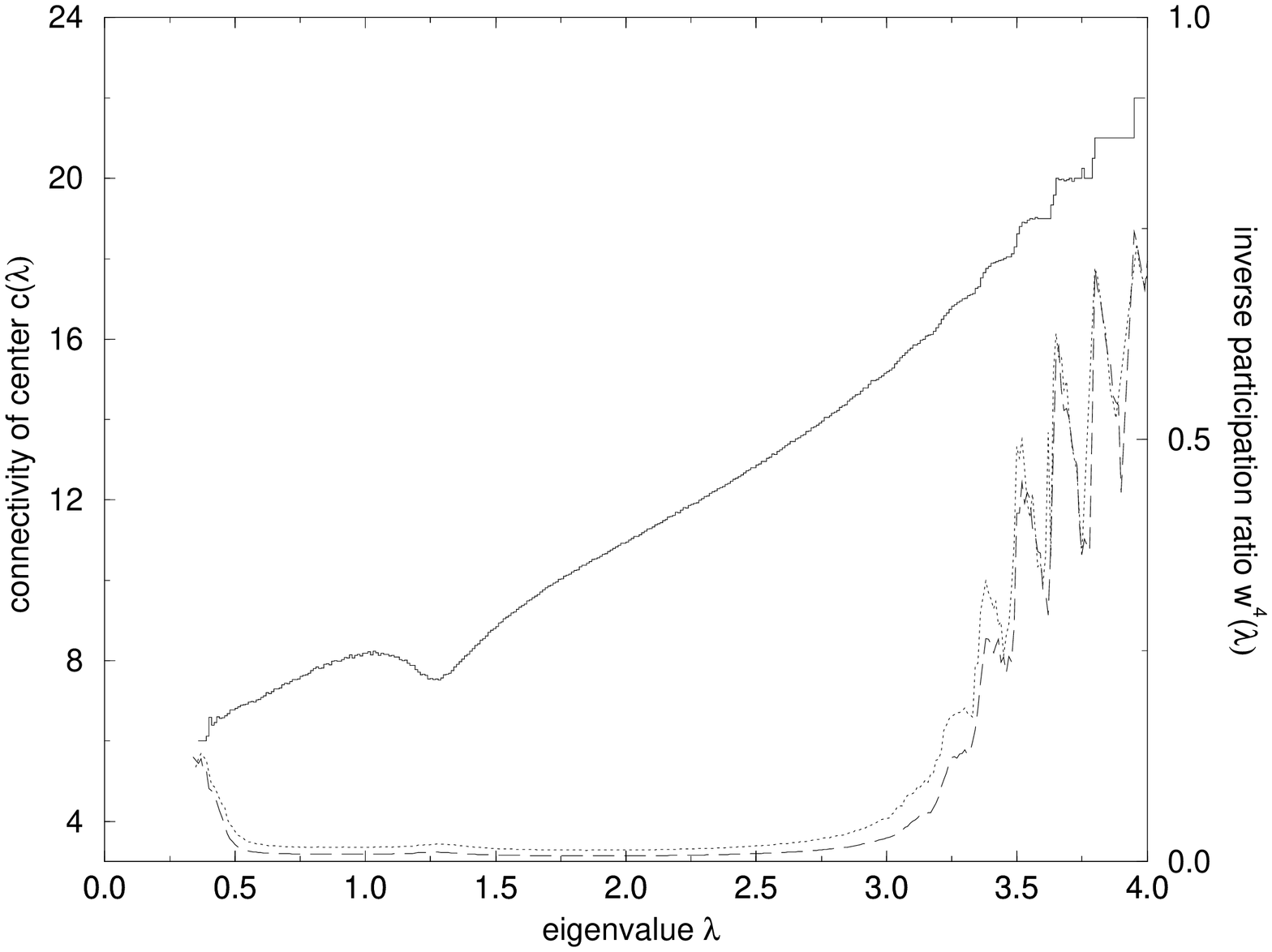}
\vskip 1cm
\caption{Results for connectivity parameters $K=3$, $q=5$.
{\bf a}~: 
density of states $p$ obtained from numerics (full curve), EMA (dotted
line) and SDA (dashed line) approximation.
{\bf b}~: inverse participation ratio $w^4$ (full curve) and center
coordination $c$ for $N=256$ (dotted curve) and $N=512$
(dashed curve) averaged over 1000
samples.   } \protect
\end{figure}

\subsection{Relations of dispersion for extended states}

Following Section II.B, we have computed the power spectra
$\overline{|\tilde w (\theta , \lambda)|^2}$ for  different
values of $\lambda $ and connectivity parameters $K=3,q=1$ and
$K=3,q=5$, see fig~4 and fig~5 respectively.
\begin{figure}
\includegraphics[width=130pt,angle=-90]{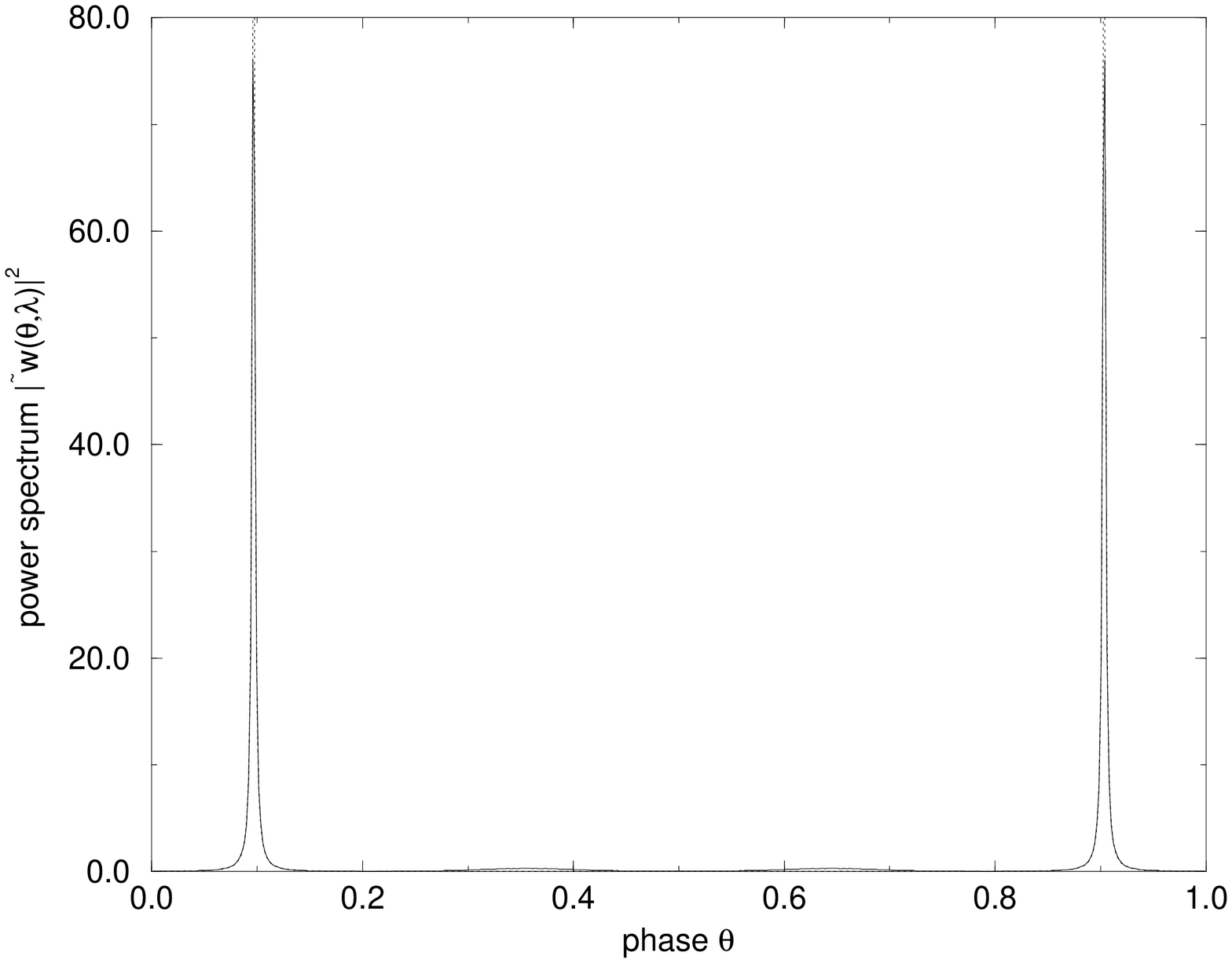}
\includegraphics[width=130pt,angle=-90]{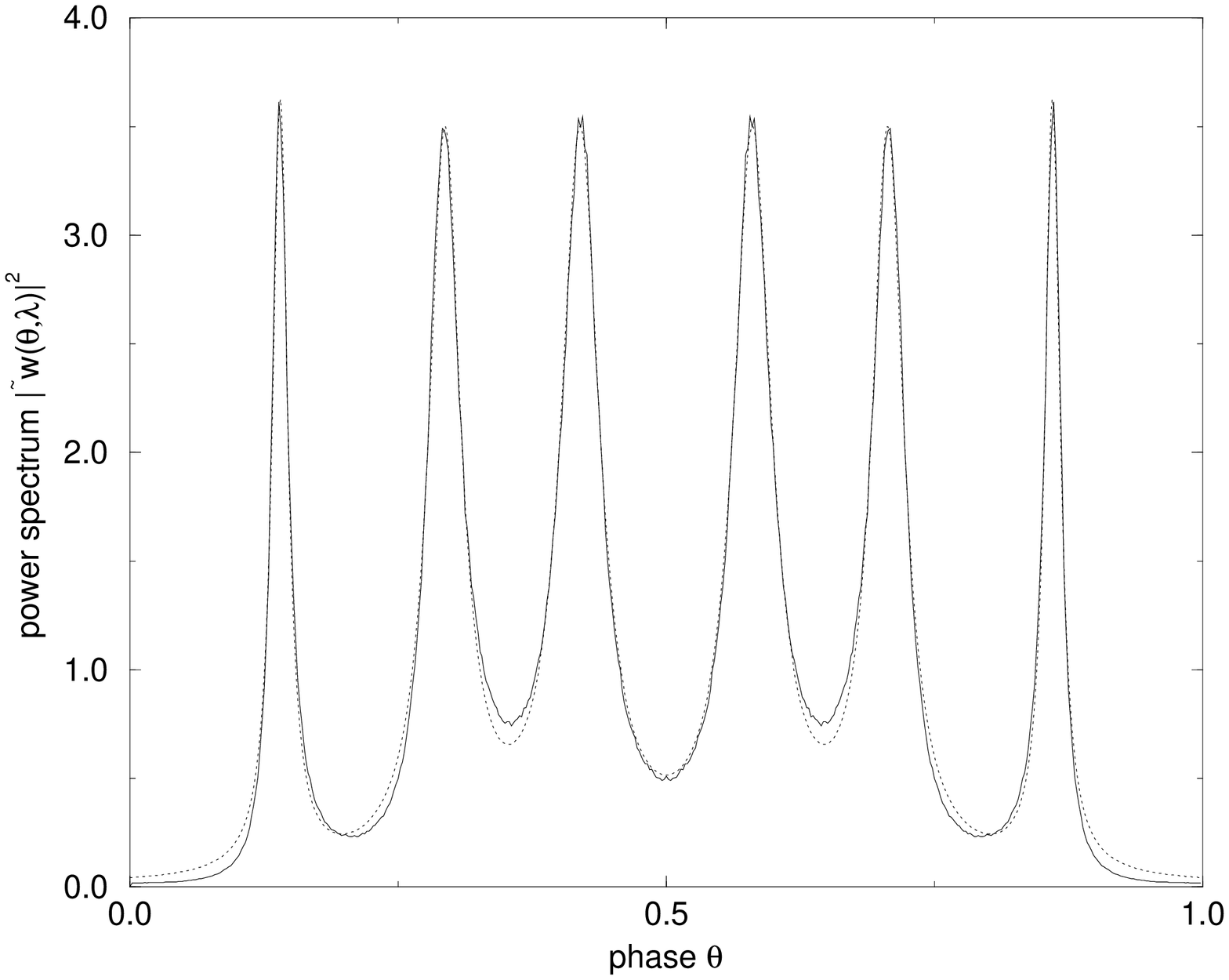}
\includegraphics[width=130pt,angle=-90]{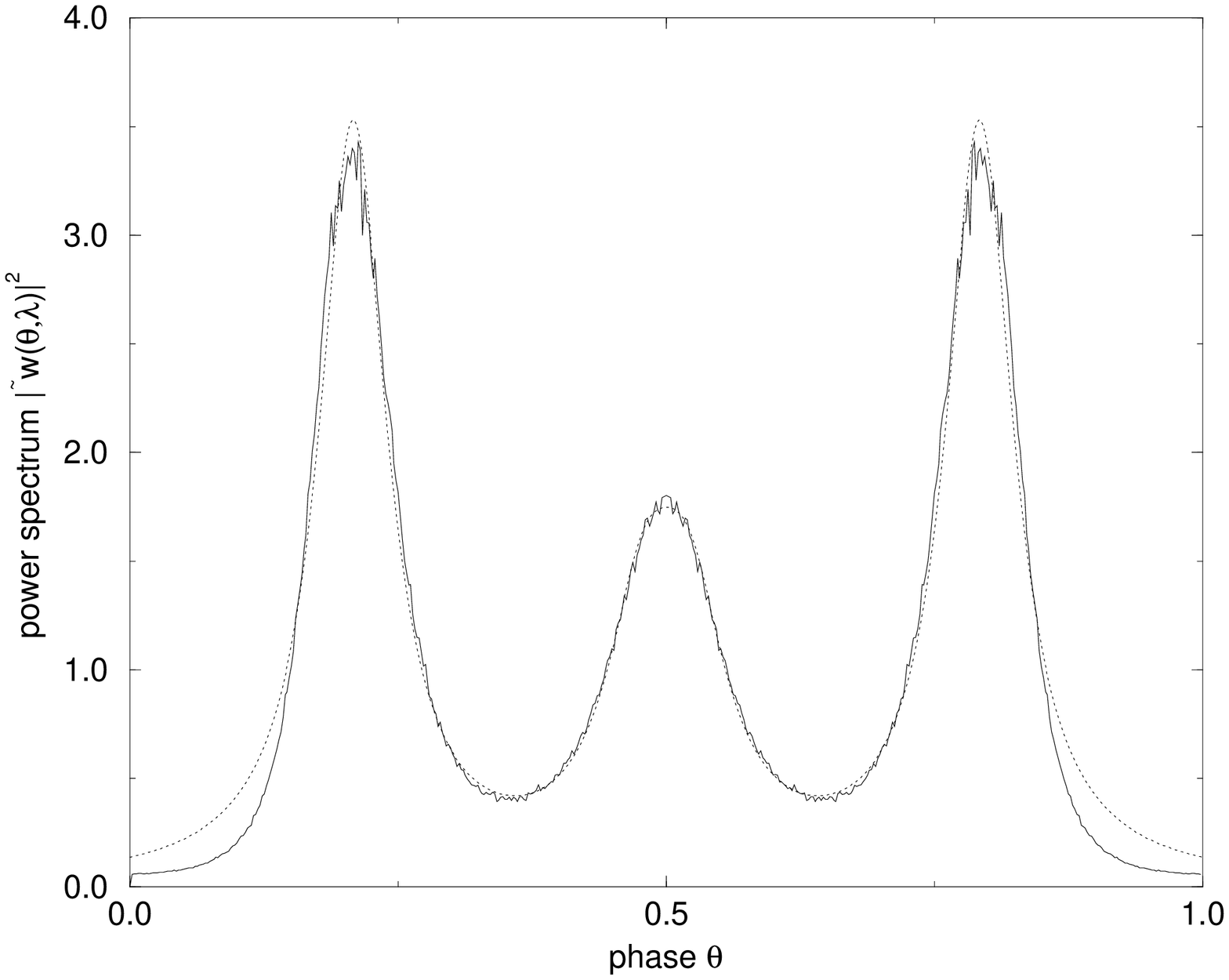}
\vskip 1cm
\caption{Power spectrum $\overline{|\tilde w (\theta ,\lambda)|^2}$ of 
the eigenvectors as a function of the phase $\theta$
for $K=3$, $q=1$ at different energy levels. The dotted lines
correspond to the Lorentzian fit (\ref{lorfit}) of Section III.B.
{\bf a}~: $\lambda = 0.8 $. {\bf b}~: $\lambda = 1.2$. {\bf c}~: 
$\lambda = 1.8$.} 
\protect
\end{figure}
Depending on the eigenvalue $\lambda$, the power spectrum is mainly 
composed of different peaks whose number varies from two to six
as in the ring case, see fig~1a. 
However, these peaks have now non-zero widths, equal to the inverse 
autocorrelation lengths of the eigenvectors. A good fit on the range
$0\le \theta \le 1$ consists in a superposition of $2\; m$ Lorentzian
spectra, see fig~4 and fig~5,
\begin{equation}
\overline{|\tilde w  (\theta , \lambda )|^2} \simeq \sum _{\ell=1} ^m \; \left[
 \frac { \tilde w_\ell }{4\pi ^2 (\theta -\phi_\ell )^2 + \sigma _\ell ^2 } +
\frac {\tilde w_\ell }{4\pi ^2(1 - \theta -\phi _\ell )^2 +\sigma _\ell ^2} 
\right] \qquad .
\label{lorfit}
\end{equation} 
Peaks are labelled so that $0 \le \phi _1 < \ldots < \phi _m $.  Close
to the pseudo-gap edge $\lambda_-$, $m$ equals two and the weight
$\tilde w_2 $ of the secondary peak gets much smaller than $\tilde
w_1$, see fig~4a and fig~5a. As a result, the estimates for $\phi_2,
\sigma_2$ becomes less accurate. Moreover, for large eigenvalues, the
Lorentzian form (\ref{lorfit}) does not fit numerical data two well. 
Consequently, the error bars on $\phi_\ell , \sigma _\ell$
increase when $\lambda$ reaches $\lambda _+$.  We have plotted fig~6
and fig~7 the wave numbers $\phi _\ell$ and the widths $\sigma _\ell$
obtained from the numerical data vs. the eigenvalue $\lambda$ for
$q=1$ and $q=5$ respectively.

At first sight, the wave numbers $\phi _\ell $ are much less affected
by disorder than the overall features of the density of states, e.g
the edges $\lambda _-$ and $\lambda _+$ and of course the inverse
lengths $\sigma _\ell$ (that are equal to zero in the ring
case). Indeed, the stationary points in the wave numbers curves fig~6a
and fig~7a, that is the merging points between $\phi _1$, $\phi_2$ and
$\phi_2$, $\phi_3$ precisely coincide with the ring values $\phi ^{\cal R}
_1$ and $\phi ^{\cal R}_2$ of Section II.C.  The main qualitative difference
between the above figures and fig~1a lies in the appearance of
a narrow area surrounded by the $\phi _2$ and $\phi _3$ lines,
centered around $\phi ^{\cal R}_2$ and extending down to $\lambda = \lambda
_-$. Thus, contrary to the ring case, the power spectrum of these
eigenstates displays more than one peak (even if the heights of the
secondary peaks are small as mentioned above).

When $\lambda$ lies in the range $\lambda _{c.o.} 
< \lambda < 1$, $\sigma _1$ gets
very small, see fig~6b and fig~7b. In presence of the disorder, these
eigenmodes have conserved an overall quasi-plane wave form. Such a
behaviour is indeed expected from Section II.E. This result explains
the quantitative agreement between the density of states of the ring
and the numerical spectrum for the small-world lattice observed in
Section III.A in the intermediate range $\lambda _{c.o.} < \lambda < 1$.

To end with, let us stress that formula (\ref{lorfit}) allows us to
define some pseudo relations of dispersion characterizing the gross
form of eigenmodes at energy $\lambda$ through some wave numbers $\phi
_\ell$ and inverse autocorrelation lengths $\sigma _\ell$.

\begin{figure}
\includegraphics[width=130pt,angle=-90]{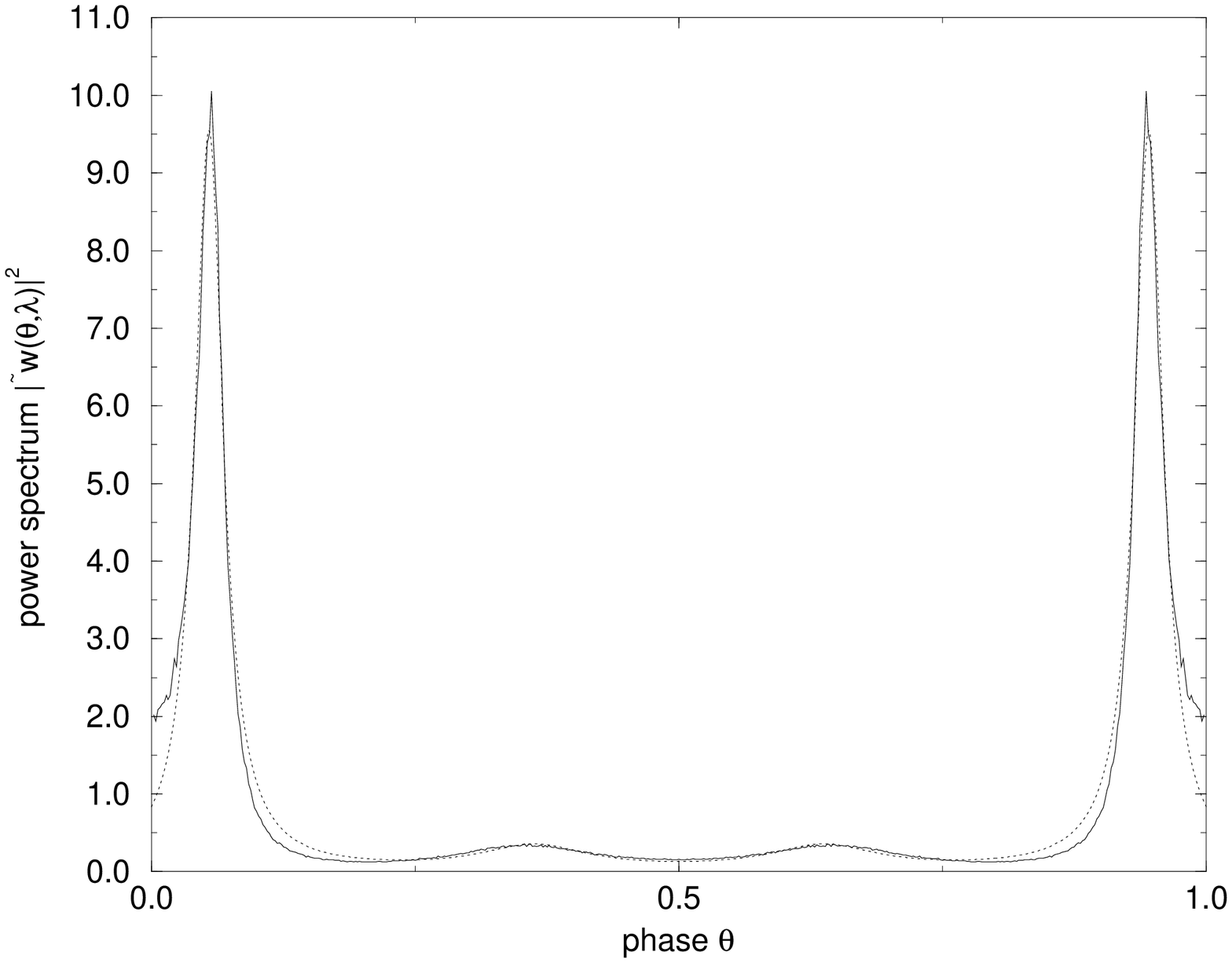}
\includegraphics[width=130pt,angle=-90]{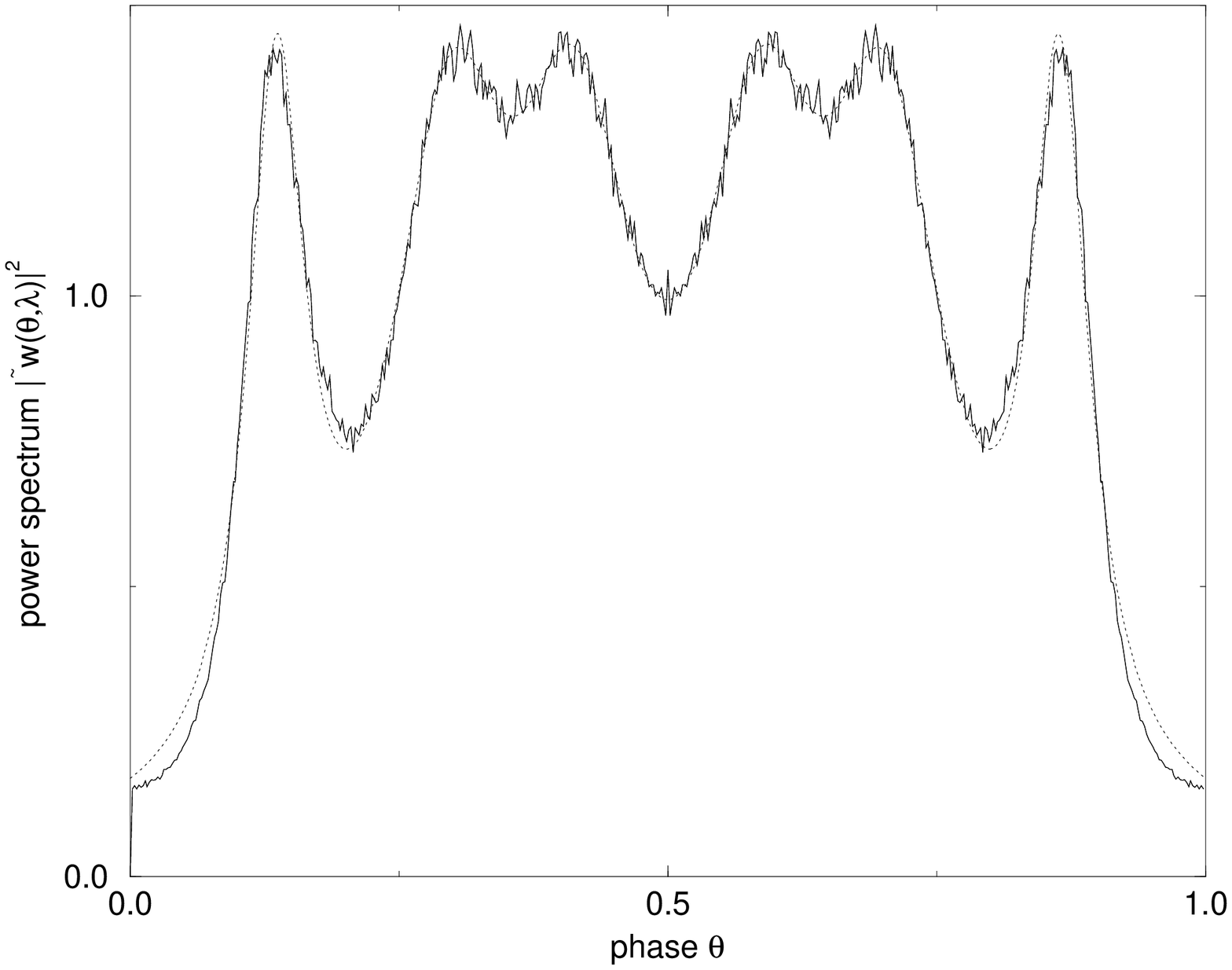}
\includegraphics[width=130pt,angle=-90]{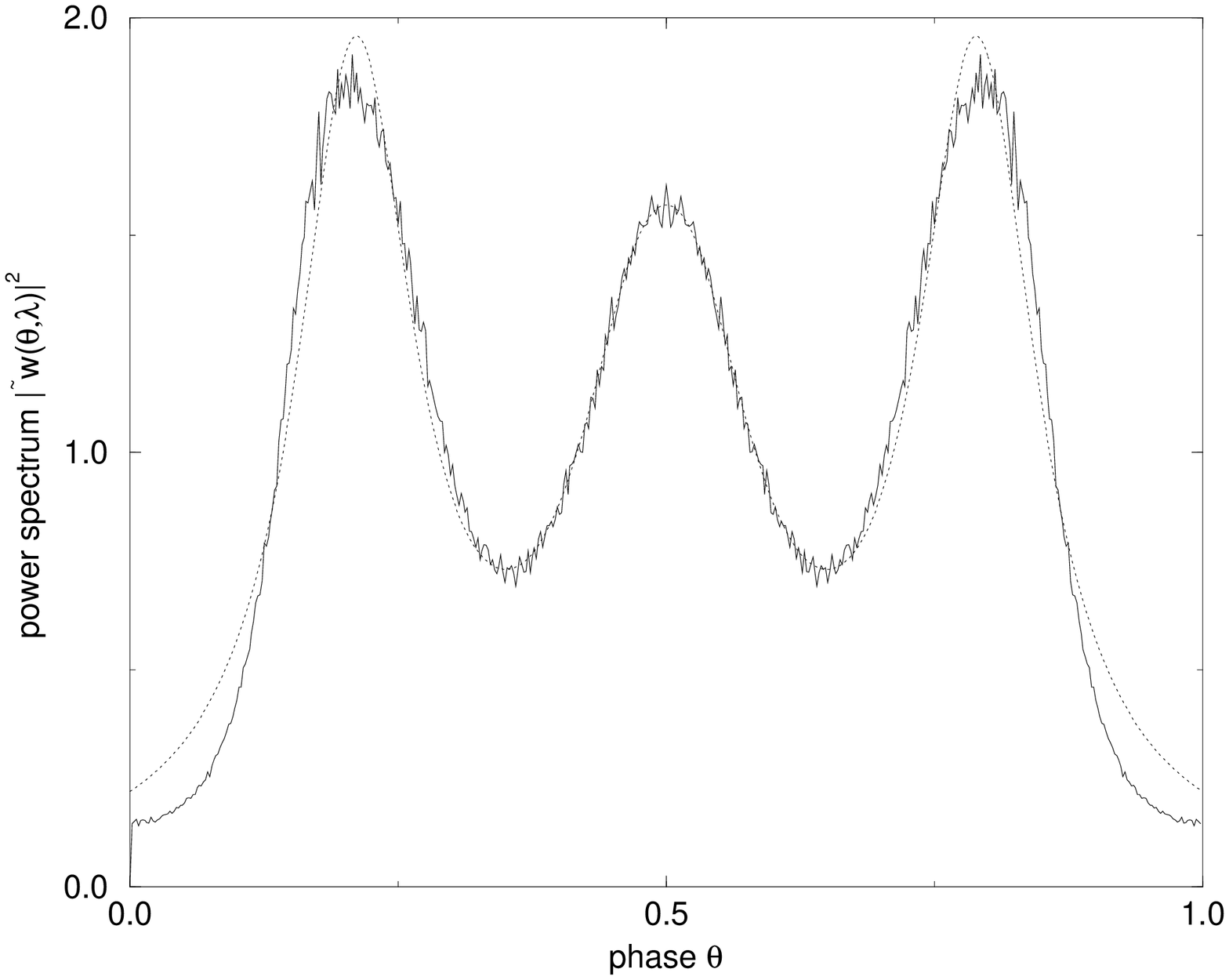}
\vskip 1cm
\caption{Same as fig~4 but with connectivity parameters $K=3$, $q=5$. 
{\bf a}~: $\lambda = 0.9$. {\bf b}~: $\lambda =1.7 $. {\bf c}~: 
$\lambda = 2.5 $.} 
\protect
\end{figure}

\begin{figure}
\vskip 1cm
\includegraphics[width=200pt,angle=-90]{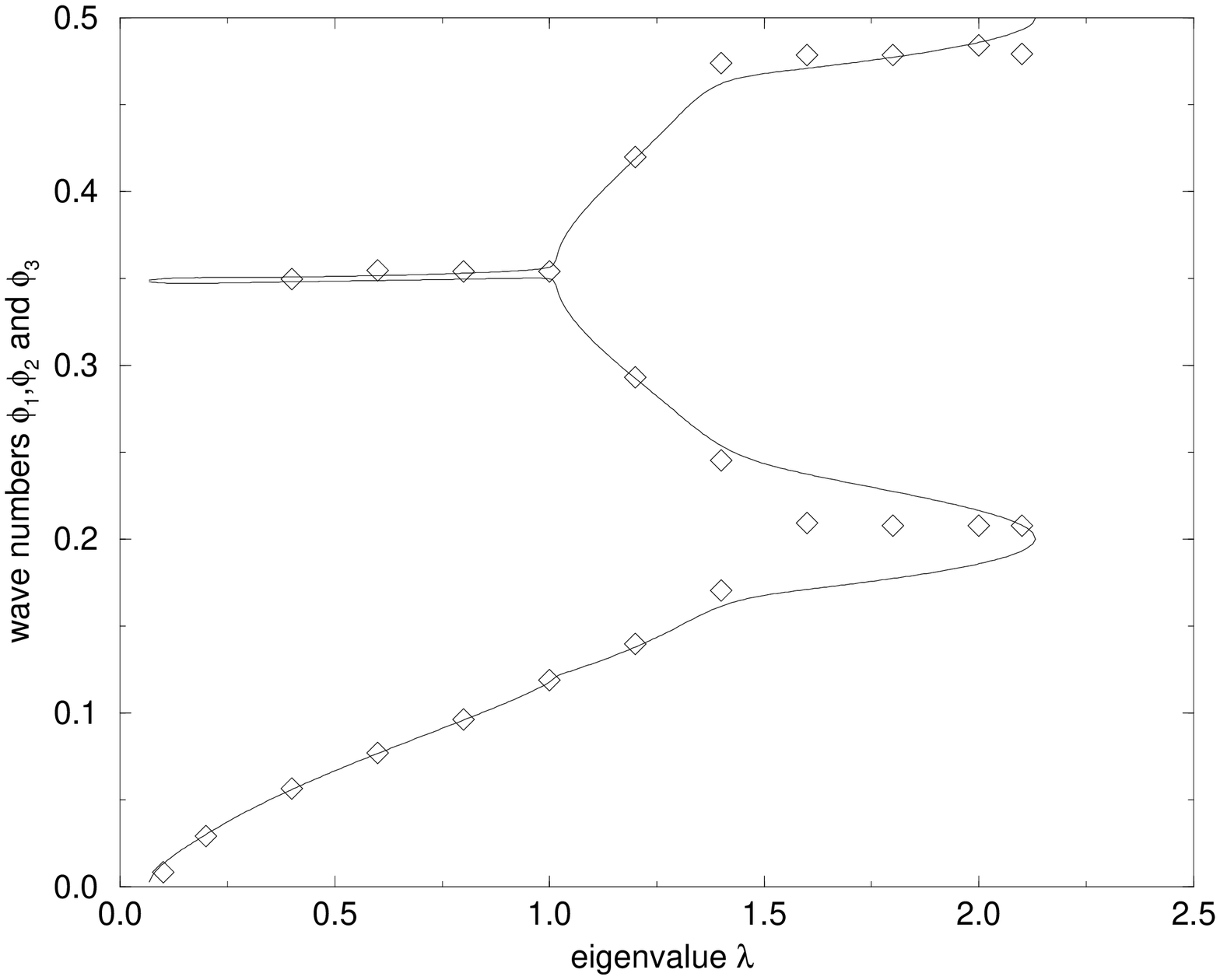}
\includegraphics[width=200pt,angle=-90]{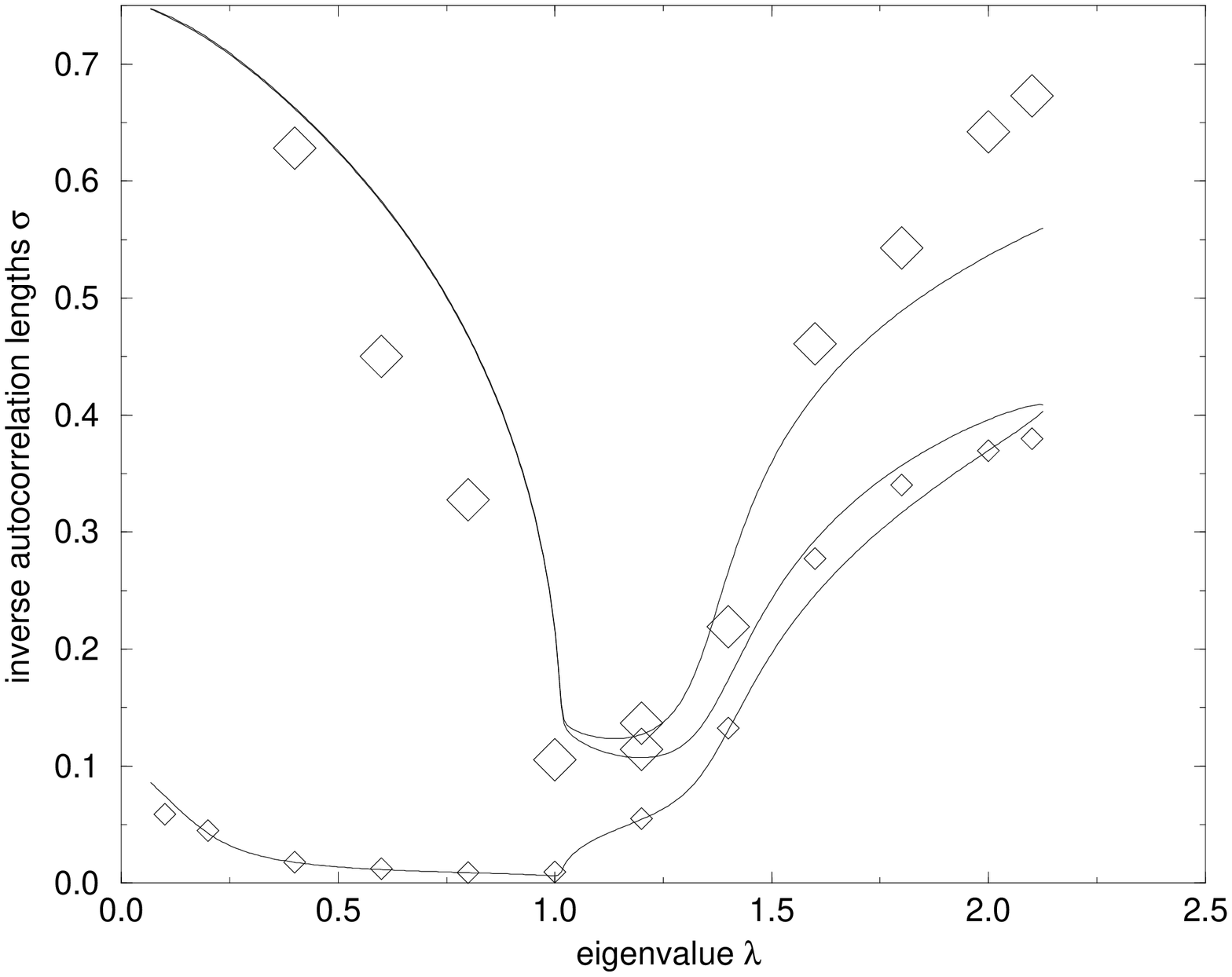}
\caption{Effective relation of dispersions for $K=3$,
$q=1$ obtained from numerical data (squares) and EMA approximation
(full line). Error bars correspond to the size of the squares. 
{\bf a}~: Wave numbers $\phi _i $, {\bf b}~: inverse
autocorrelation lengths $\sigma _i$.}
\protect
\end{figure}

\begin{figure}
\includegraphics[width=200pt,angle=-90]{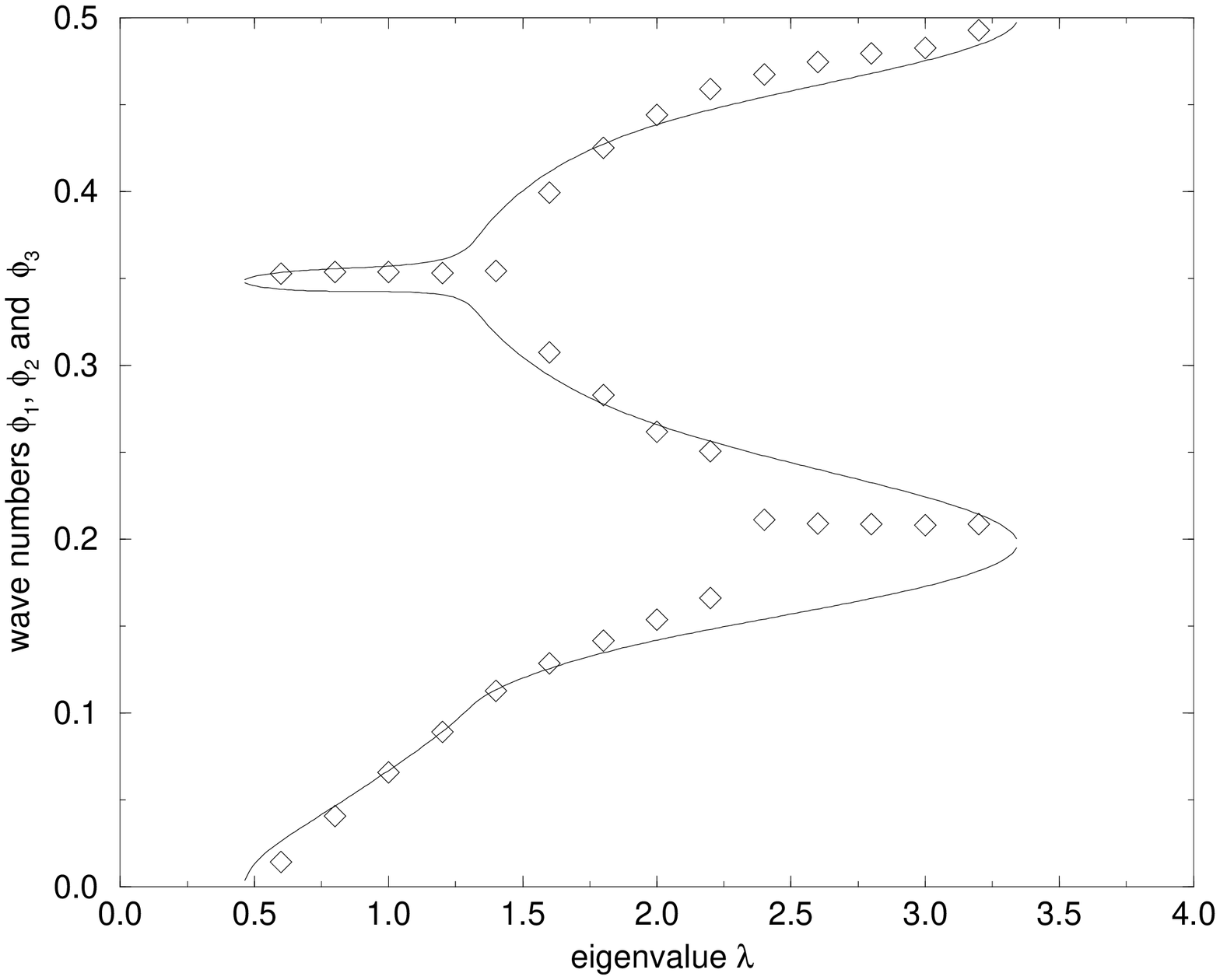}
\includegraphics[width=200pt,angle=-90]{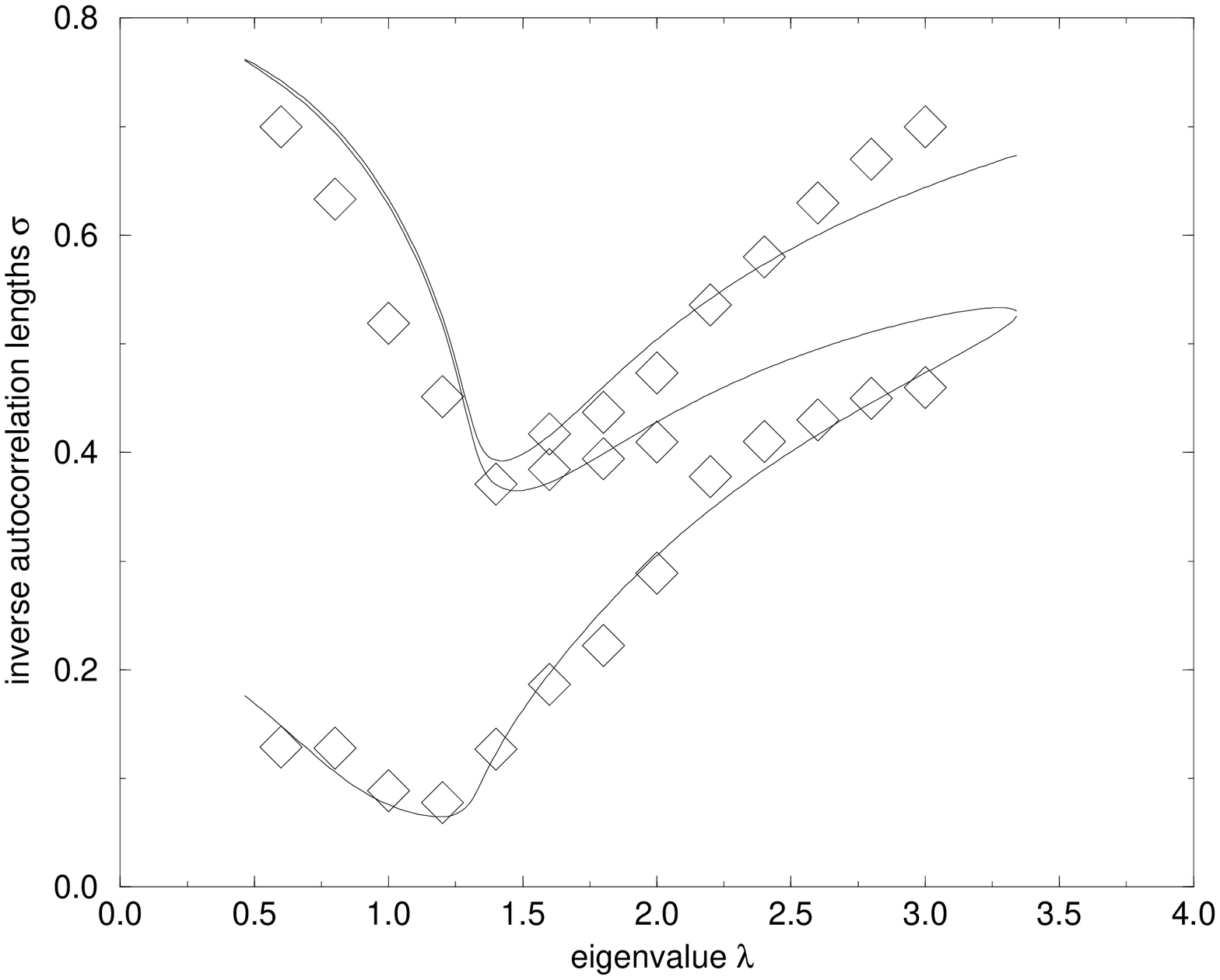}
\vskip 1cm
\caption{Same as fig~6 but with connectivity parameters $K=3$, $q=5$.}
\protect
\end{figure}

\subsection{Localized eigenstates}

Following Section II.D, we have measured for each eigenvector $w _{i}$
the connectivity $c_\ell $ of its center, that is the site $i_0$ with
maximum component $|w _{i_0 , e}|$ \cite{sda}. The mean connectivity
$c(\lambda )$ of the centers of eigenvectors having eigenvalue
$\lambda$ is plotted fig.~2b. It is a smooth monotonous function of
$\lambda$ in the central part of the spectrum.  In the localized
region, $c(\lambda )$ is constant over a given peak and integer-valued
(for $c\ge c_+ = 10$ on the right side); the center connectivity
abruptly jumps when $\lambda$ crosses the borders between peaks.

Table~I and II list the weights $p_c^{NUM}$ of the localized peaks related
to connectivities $c$, that is the integrated densities of eigenvalues
belonging to them.  The corresponding eigenvalues $\lambda _c ^{NUM}$
are also given. They are measured at the top of the peaks with
absolute error $\pm 0.005$ whereas the relative error on $p_c^{NUM}$
is about $20\%$. The reliability of numerical results suffers from two
effects. First, as $\lambda $ decreases, localized peaks get closer
and closer to the flank of the extended spectrum and the
determination of the top of the peaks $\lambda _c ^{NUM}$ becomes less
and less accurate. Secondly, for large eigenvalues, the corresponding
statistical events are rare and fluctuate drastically from sample to
sample leading to a poor accuracy on $p_c^{NUM}$.  Note however that
$p_c^{NUM}$ is of the same order of magnitude (but smaller as it
should be due to the screening effect exposed in \cite{sda}) as the
fraction of sites having $c=2K+c'$ neighbors, with $c'$ given by a
Poisson law of parameter $q$.

Contrary to the random graph case \cite{sda}, localized peaks are not
found on the left side of the spectrum. Indeed, the minimal
connectivity of a site in the small-world case is $c=2K$ and defects
with low connectivities with respect to the average coordination
$2K+q$ can be seen at very strong disorder only.

\begin{table}
{\footnotesize
$$
\begin{array}{||c|c|c|c|c|c|c|c|c||}
\hline \hline c  & 11 & 12 & 13 & 14  \\ \hline \hline
\lambda ^{NUM}   & 2.175 & 2.315 & 2.460 &  2.615 \\ 
\hline \lambda ^{SDA} & 2.195 & 2.325 & 2.469 & 2.619  \\ \hline \hline
p ^{NUM}  & 0.60  & 0.78  & 0.90 & 0.44 \\ 
\hline p ^{SDA}  & 0.550 & 0.686 & 0.755 & 0.798
\\ \hline \hline 
\end{array}
$$
}
\caption{Case $K=3,q=1$~: 
Weights $p $ ({\em i.e.} integrated density of eigenvalues
belonging) of the localized peaks (divided by the Poisson factor
$e^{-q} q^{c'} / c'!$ with $c'=c-2K$) and corresponding 
eigenvalues $\lambda$ for defect connectivities $c$. 
Results are obtained from numerical simulations (NUM) and SDA. The
accuracy of numerical values is discussed in Section III.C. }
\end{table}

\begin{table}
{\footnotesize
$$
\begin{array}{||c|c|c|c|c|c|c|c|c||}
\hline \hline c & 18 & 19 & 20 & 21 & 22 \\ \hline \hline
\lambda ^{NUM} & 3.425 & 3.545 & 3.685 & 3.850 & 3.990  \\ \hline 
\lambda ^{SDA} & 3.433 & 3.556 & 3.696 & 3.841 & 3.990  \\ \hline \hline
p ^{NUM} & 0.60 & 0.62 & 0.79 & 0.77 & 0.60 \\ \hline 
p ^{SDA} &  0.574 & 0.696  & 0.762 & 0.802  & 0.831
\\ \hline \hline 
\end{array}
$$
}
\caption{Same as Table~I but with connectivity parameters $K=3$, $q=5$.}
\end{table}

%
%

\section{Theoretical framework}

\subsection{The average over the disorder}

Once averaged over the disorded, the resolvent (\ref{fieldd}) can be 
written as the propagator of a replicated Gaussian field theory 
\cite{thou}
\begin{eqnarray}
\overline{G^{\cal S} _{jk}(\lambda + i \epsilon ) } &=& \lim_{n\rightarrow 0}
- \frac{i}{n}  \int \prod _{i} d\vec x _{i} 
(\vec x _j .\vec x _k )\; \overline{ 
\exp \left(\frac{i}{2}( \lambda + i\epsilon )
\sum _i \vec x _{i} ^{\; 2} + \frac{i}{2} \sum _{i<\ell} W^{\cal S}_{i\ell} 
(\vec {x}_{i} - \vec {x}_{\ell} )^2 \right)}
 \quad
\label{field}
\end{eqnarray}
Replicated fields $\vec x _i$ are $n$-dimensional vector fields
attached to each site $i$. 

The Laplacian on the ring ${\cal R}$ is non fluctuating and we have to
perform the average over the random graph ${\cal G}$ only. Due to the
building process described in Section II.A, the averages over
different pairs of points are decoupled and we find
\begin{eqnarray}
\overline{
\exp \left( \frac{i}{2} \sum _{i<\ell} W^{\cal G} 
_{i\ell}  (\vec {x}_{i} - \vec {x}_{\ell} )^2 \right)}
&=& \prod _{i < \ell} \left[ 1 - \frac qN + \frac qN
\exp \left( -\frac{i}{4K} (\vec {x}_{i} - \vec {x}_{\ell} )^2 \right)
\right] \nonumber \\
&\simeq & \exp \left[ - \frac {qN}2 + \frac qN \sum _{i<\ell} 
\exp \left( -\frac{i}{4K} (\vec {x}_{i} - \vec {x}_{\ell} )^2 \right)
 \right] \quad ,
\label{field1}
\end{eqnarray}
to largest order in $N$. 
An important property of the last term in (\ref{field1}) is that all
pair of sites $i,\ell$ interact together with the same coupling. 
Therefore, introducing the
density $\rho (\vec x )$ of replicated vectors $\vec x _i$ equal to
$\vec x$,
\begin{equation}
\rho (\vec x ) = \frac 1N \sum _{i=1}^N \; \prod _{a=1}^n \delta (
x^a - x_i^a ) \qquad ,
\label{rho}
\end{equation}
by means of a functional Lagrange multiplier $\omega (\vec x )$, we
obtain
\begin{eqnarray}
\overline{
\exp \left( \frac{i}{2} \sum _{i<\ell} W^{\cal G} 
_{i\ell}  (\vec {x}_{i} - \vec {x}_{\ell} )^2 \right)}
= \int {\cal D}\rho (\vec x ) {\cal D}\omega (\vec x )
\; \exp&& \left[ - \frac {iN}2 \int d\vec x \; \omega (\vec x) \rho ( \vec x)
+ \frac i2 \sum _{i=1}^N  \omega (\vec x _i ) \right.-\frac {qN}2 
\nonumber \\
&& \left. + \frac {qN}2 \int d\vec x d\vec y \; \rho ( \vec x)
\rho ( \vec y)\;
\exp \left( -\frac{i}{4K} (\vec {x} - \vec y )^2 \right)
 \right] \quad ,
\label{field2}
\end{eqnarray}
where the above expression includes a double path-integral over $\rho$
and $\omega$ functions. We shall come back in Section IV.C to the
significance of the latters. 

We can now take into account the contributions coming from the
diagonal term and from $W^{\cal R}$ in (\ref{field}). The $n^{th}$
moment of the partition function $Z$ of the replicated Gaussian field
theory introduced in (\ref{field})
\begin{equation}
\overline{Z^n} \equiv \int \prod _{i} d\vec x _{i} \; \overline{ 
\exp \left(\frac{i}{2}( \lambda + i\epsilon )
\sum _i \vec x _{i} ^{\; 2} + \frac{i}{2} \sum _{i<\ell} W^{\cal S}_{i\ell} 
(\vec {x}_{i} - \vec {x}_{\ell} )^2 \right)}
\label{zn1}
\end{equation}
may be written as
\begin{equation}
\overline{Z^n}  = \int {\cal D}\rho (\vec x ) {\cal D}\omega (\vec x )
\; \exp \left( - N {\cal F} (\rho ,\omega ) \right) \qquad ,
\label{zn2}
\end{equation}
with
\begin{equation}
{\cal F} (\rho ,\omega ) = \frac i2 \int d\vec x \; \omega (\vec x) 
\rho ( \vec x)-  \frac q2
\int d\vec x d\vec y \; \rho ( \vec x) \rho ( \vec y)\;\left[
\exp \left( -\frac{i}{4K} (\vec {x} - \vec y )^2 \right) -1
\right] -\frac 1K
\log \Lambda _M (\omega ) \quad .
\label{field3}
\end{equation}
In the above expression, $ \Lambda _M (\omega )$ denotes the largest
eigenvalue of a transfert matrix $T$ we shall define in next section
and which depends on the function $\omega$.
Note that (\ref{field3}) is exact if $N$ is a multiple of $K$. 

The analogy with the theory of polymers initiated
by Edwards \cite{deg,edw} is clear once the $\rho$ variables have been 
integrated out in (\ref{zn2}). We are then left with a one-dimensional
model of free-energy density $- \log \Lambda _M (\omega ) /K$ 
in presence of a random potential $\omega$ distributed according to a
Gaussian law. For the small-world lattice, this is no approximation
since the virial expansion includes a unique corrective term to the
non-interacting case, see Section IV.C.

\subsection{Transfer matrix}

Each element of the transfer matrix $T$ arising in the computation of 
the free-energy functional (\ref{field3}) is labelled by $2K$
replicated vectors $\vec x_1 , \vec x_2 , \ldots ,\vec x_K$ and 
$\vec y_1 , \vec y_2 , \ldots ,\vec y_K$. From (\ref{field}) and
(\ref{field1}), $T$ reads
\begin{eqnarray}
T[ \vec x_1 , \vec x_2 , \ldots ,\vec x_K &;& \vec y_1 , \vec y_2 ,
\ldots ,\vec y_K ] = \exp \left\{ \frac i4 (\lambda +i \epsilon )
\sum _{j=1}^K [\vec x_j ^{\; 2} + \vec y_j ^{\; 2} ] + \frac i4 \sum _{j=1}^K 
[ \omega (\vec x_j  )+\omega (\vec y_j  ) ] \right. \nonumber \\
&& \left. - \frac i{8K}
\sum _{1\le j<\ell \le K } [ (\vec x_j -\vec x_\ell )^2 + 
(\vec y_j -\vec y_\ell )^2 ] -\frac i{4K} \sum _{ j=1}^K \sum
_{\ell =1} ^j (\vec x_j -\vec y_\ell )^2 \right\} \qquad .
\label{transf}
\end{eqnarray}
Due to the last term in (\ref{transf}), $T$ is not a symmetric
matrix. However, defining its ``mirror'' matrix $T^{\#}$  through
\begin{equation}
T ^{\#}[ \vec x_1 , \vec x_2 , \ldots ,\vec x_K ; \vec y_1 , \vec y_2 ,
\ldots ,\vec y_K ] =
T[ \vec y_K , \vec y_{K-1} , \ldots ,\vec y_1 ; \vec x_1 , \vec x_2 ,
\ldots ,\vec x_K ] \qquad ,
\label{mirror1}
\end{equation}
it appears from (\ref{transf}) that
$T=T^{\#}$. 
Let us now define the mirror $v^\#$ of a vector $v$ through
\begin{equation}
v^\# [\vec y_1 , \vec y_2 ,
\ldots ,\vec y_K ] = v [\vec y_K , \vec y_{K-1} , \ldots ,\vec y_1]
\quad .
\end{equation}
Then, it can be easily checked that 
the left eigenvectors of $T$ are the mirrors of its right eigenvectors. 
Consequently, in the large $N$ limit, the matrix entries of $T^N$
simplify to 
\begin{equation}
(T^N) [ \vec x_1 , \vec x_2 , \ldots ,\vec x_K ; \vec y_1 , \vec y_2 ,
\ldots ,\vec y_K ] \simeq  (\Lambda _M) ^N \; 
v^\# _M [\vec y_1 , \vec y_2 , \ldots ,\vec y_K ] v_M [\vec x_1 , \vec x_2 ,
\ldots ,\vec x_K ]
\quad , \label{projec}
\end{equation}
where $v_M$ is the maximal eigenvector of $T$ associated to $\Lambda _M$.

\subsection{Meaning of the order parameters $\rho$ and $\omega$}

In the large $N$ limit, the $n^{th}$ moment of the partition function
(\ref{zn1},\ref{zn2}) can be calculated using the saddle-point method. The 
optimization of the free-energy functional ${\cal F}$ (\ref{field3})
gives the values of  $\rho (\vec x )$  and $\omega (\vec x)$, as well
as $v_M [\vec x_1 , \vec x_2 , \ldots ,\vec x_K ]$ at saddle-point.

The significance of $\rho$ is obvious from definition
(\ref{rho}); this is the probability distribution of replicated
vectors $\vec x_i$. Therefore, $\rho$ gives access to the average
resolvent (\ref{field})
\begin{equation}
\overline{ G^{\cal S}_{jj} (\lambda + i\epsilon )} = \lim_{n\rightarrow 0}
- \frac in \int d\vec x \; \rho (\vec x ) \; \vec x ^{\; 2} \qquad ,
\label{Grho}
\end{equation}
and to the density of states through (\ref{spectre}).

More generally, within the transfer matrix formalism developed in
Section IV.B, the joint probability distribution of  replicated
vectors $\vec x_i , \vec x_{i+1} , \ldots , \vec x_{i+K-1}$ is related
to the maximal eigenvector $v_M$ of $T$. Normalizing $v_M$ to unity,
 this joint probability reads (\ref{projec})
\begin{equation}
\rho _{joint} ( \vec x_i , \vec x_{i+1} , \ldots , \vec x_{i+K-1} )
= v_M [ \vec x_i , \vec x_{i+1} , \ldots , \vec x_{i+K-1}] \;
v_M^\# [ \vec x_i , \vec x_{i+1} , \ldots , \vec x_{i+K-1}]
\quad .
\label{joint}
\end{equation}
In particular, the one point density $\rho$ is found back when all but
one vectors are integrated out in the above formula
\begin{equation}
\rho (\vec x ) = \int d\vec x_2 \ldots d\vec x_K \;
v_M [ \vec x , \vec x_{2} , \ldots , \vec x_{K} ] \;
v_M^\# [ \vec x , \vec x_{2} , \ldots , \vec x_{K} ]
\qquad .
\label{rho1}
\end{equation}
Identity (\ref{rho1}) is precisely the stationary condition of ${\cal
F}$ (\ref{field3}) with respect to $\omega$. 

The meaning of $\omega$ is less straightforward but emerges from the 
extremization condition of ${\cal F}$ with respect to $\rho$,
\begin{equation}
\frac i2 \omega (\vec x ) = q\; \int d\vec y \; \rho ( \vec y)\;\left[
\exp \left( -\frac{i}{4K} (\vec {x} - \vec y )^2 \right) -1
\right]
\qquad .
\label{omega1}
\end{equation}
The above equation is strongly reminiscent of the first correction to
the logarithm of the density obtained from Mayer's expansion of
interacting gas\cite{Morita}. Indeed, this close relationship has been
exposed in \cite{sda} (see also \cite{remi} for the similar case of
finite-connectivity spin-glasses). The replicated field $\vec x_i$
attached to site $i$ may be interpreted as the position of particle $i$
in a $n-$dimensional abstract space. Particles see an external
harmonic potential $-\frac i2 (\lambda + i \epsilon )
\vec x ^{\; 2}$ and interact with each other through a
two-body quadratic energy $-\frac i2 W^{\cal S}_{ij} (\vec {x _i} - \vec x_j )^2$, see
(\ref{zn1}).  In the absence of any $W^{\cal S}$ matrix,
the density of a non-interacting gas is recovered
\begin{equation}
\rho _{0} (\vec x ) = \exp \left( \frac i2 (\lambda + i \epsilon )
\vec x ^{\; 2} \right) \qquad ,
\label{pg}
\end{equation}
obtained from equation (\ref{zn1}). The first virial coefficient, that
is the first correction to $\log \rho (\vec x )$ in the expansion in
powers of $\rho $ (and not in powers of the fugacity $\rho _{0}$)
is precisely given by the r.h.s of (\ref{omega1})
\cite{sda,Morita}. It turns out that Mayer's expansion for random
graph do not include higher order terms, as is expected from the
similarity between random graph and trees.

Therefore, $- \frac i2 \omega ( \vec x ) $ may be understood as the
effective potential due to interactions with other particles seen by 
a particle located in $\vec x$. This
interpretation will be useful in deriving the single defect
approximation of Section VI.

\subsection{Autocorrelation function of the eigenvectors}

We have seen in Section II.B that the autocorrelation functions of
eigenvectors are related to the resolvent $\overline{G^{\cal S}_{jk}}$,
see identity (\ref{rel1}). By definition, the latter is the
average scalar product of replicated vectors $\vec x_j$ and $\vec x_k$
(\ref{field}) with the Gaussian measure (\ref{zn1}).  Within the
transfer matrix formalism we have introduced in Sections IV.B and
IV.C, this mean dot product can be computed from the knowledge of all
eigenvectors $v_\ell$ and eigenvalues $\Lambda _\ell$ of $T$. We
decompose the distance $d=|j-k|$ between sites $j$ and $k$ as $d= d_0 + K \;
d_1$ with $0\le d_0 \le K-1$ and $d_1$ integer-valued. 
Then the average resolvent reads
\begin{equation}
\overline{G^{\cal S}_{jk}(\lambda + i\epsilon )}
 = \lim _{n\to 0} - i \sum _{\ell = 1}^\infty \langle v_M ^\# . X _1 ^1
. v_\ell \rangle \; \langle v_\ell ^\# . X _{1+d_0} ^1
. v_M \rangle \; \left( \frac{\Lambda _\ell}{\Lambda _M} \right)^{d_1}
\quad . \label{corre}
\end{equation}
In the above equation, the maximal eigenvalue of $T$ coincides with
$\ell =1$  ($\Lambda _1 \equiv \Lambda _M$) and increasingly excited states
correspond to $\ell \ge 2$. The $X ^a _b$ operator measures the value
of the $a^{th}$ component of the replicated field $\vec
x$ appearing at the $b^{th}$ place in $v_\ell$ (with $1\le b \le K-1$)
and  $\langle v_M^\# . X . v_\ell \rangle$ denotes the matrix element
of $X^a_b$ between states $v_M$ and $v_\ell$,
\begin{equation}
\langle v_M^\# . X ^a _b . v_\ell \rangle = \int d\vec x_1 \ldots d\vec x_K \;
x^a_b \; v_M ^\#[ \vec x_1 , \vec x_{2} , \ldots , \vec x_{K} ] \;
v_\ell [ \vec x_1 , \vec x_{2} , \ldots , \vec x_{K} ]
\qquad . \label{lan}
\end{equation} 
For small distances $d\le K-1$, $d_1=0$ and the sum over eigenstates
in formula (\ref{corre}) can be carried out to obtain
\begin{equation}
\overline{G^{\cal S}_{jk}(\lambda + i\epsilon )}
 = \lim _{n\to 0} - i\ \langle v_M ^\# . X _1 ^1 \; X _{1+d} ^1 . v_M \rangle 
, \qquad 0 \le d \le K-1 . 
\end{equation}
Using identity (\ref{rho1}), the above equation gives back the $d=0$
resolvent (\ref{Grho}). 

Expression (\ref{corre}) may also be used to compute the
autocorrelation function of eigenvectors at large distance $d$, that
is when the sum in (\ref{corre}) is dominated by the $\ell=2$
contribution. The ratio $\tau =\Lambda _2 /\Lambda _M$ whose
modulus is by definition smaller than (or equal to) unity can be
written as 
\begin{equation}
\tau = \exp \left( -  K \sigma + i 2\pi K
\phi \right) \qquad , \label{rel3}
\end{equation}
with $\sigma \ge 0$. Therefore, at large distances $d$, 
the average resolvent scales as
\begin{equation}
\overline{G^{\cal S}_{jk} (\lambda + i\epsilon )} \sim \exp 
(- \sigma \; d + i 2\pi \phi \; d ) \qquad ,
\label{rel2}
\end{equation}
where both $\sigma$ and $\phi$ depend on the energy level $\lambda$.
Consequently, $\sigma$ may be seen as the inverse autocorrelation
length of the eigenmodes associated to level $\lambda$ and $\phi$ to
the typical wave number of the latters. We shall explicitely compute
the inverse length $\sigma$ and wave number $\phi$ in Section V.D
within the effective medium approximation and compare them to the numerical
results of Section III.B.

%
%

\section{Effective medium approximation}

The exact extremization of free-energy function (\ref{field3}) is
an awkward task. We start by solving this problem using an
effective medium approximation (EMA) which gives a good description of
the extended part of the spectrum.

\subsection{Presentation of the approximation}

The starting point of EMA is the exact expression of the density
$\rho$ obtained from (\ref{zn1}) \cite{sda},
\begin{equation}
\rho (\vec x ) = \frac 1N \overline{ \sum _{i=1}^N C_i \exp \left(
\frac{i\; \vec x ^{\; 2}}{2 [(\lambda + i\epsilon )
1-W^{\cal S}]^{-1} _{ii}} \right)}  \quad,
\label{inter}
\end{equation}
where the $C_i$ are normalization constants going to unity as $n$ vanishes.

In the extended part of the spectrum, we expect all matrix elements
appearing in (\ref{inter}) to be of the same order of magnitude and
thus $\rho(\vec x )$ to be roughly Gaussian. We therefore assume the
following Ansatz for the density
\begin{equation}
\rho ^{EMA} (\vec x ) = \left( 2\pi i g^{EMA} \right)^ {-\frac{n}{2}}\exp
\left( \frac{i \; \vec x ^{\; 2}}{2\; g^{EMA}} \right) \quad .
\label{ema}
\end{equation}
In addition, the effective potential $\omega$ we choose to be
harmonic, see Section IV.C,
\begin{equation}
\omega ^{EMA} (\vec x ) = \hat g^{EMA} \; \vec x ^{\; 2} \quad .
\label{ema2}
\end{equation}
EMA is therefore implemented by inserting the Gaussian Ansatz
(\ref{ema},\ref{ema2}) into functional ${\cal F}$. Due to the choice
(\ref{ema2}) for $\omega$, the transfer matrix $T$ simplifies to the
tensorial product of $n$ identical matrices
\begin{eqnarray}
T ^{EMA} [ x_1 , x_2 , \ldots , x_K ;  y_1 ,  y_2 ,
\ldots ,y_K ] &=& \exp \left\{ \frac i4 (\hat g^{EMA} + \lambda +i \epsilon )
\sum _{j=1}^K [x_j ^{\; 2} +  y_j ^{\; 2} ] \right. \nonumber \\
&& \left. - \frac i{8K}
\sum _{1\le j<\ell \le K } [ (x_j - x_\ell )^2 + 
(y_j - y_\ell )^2 ] -\frac i{4K} \sum _{ j=1}^K \sum
_{\ell =1} ^j (x_j - y_\ell )^2 \right\} \qquad .
\label{transfema}
\end{eqnarray}
The resulting free-energy per replica reads in the limit $n\to 0$,
\begin{equation}
{\cal F}^{EMA} (g^{EMA},\hat g^{EMA}) = -\frac 12 g^{EMA} \hat g^{EMA} + \frac q4 \log \left( 
1 - \frac {g^{EMA}}K \right) - \frac 1K \log \Lambda ^{EMA}_M(\hat g^{EMA})
\qquad ,
\label{ema3}
\end{equation}
where $\Lambda ^{EMA}_M(\hat g^{EMA})$ is the largest eigenvalue of $T^{EMA}$
(\ref{transfema}). 

\subsection{Self-consistent equations}

To compute $\Lambda ^{EMA}_M(\hat g^{EMA})$, we take advantage of
the Gaussian structure of $T^{EMA}$ and look for a maximal eigenvector
of the form
\begin{equation}
v^{EMA}_M [ x_1 ,  x_{2} , \ldots ,  x_{K} ] = v_0. \exp \left(
\frac i2 \sum _{q,r=1}^K V _{qr}  \; x_q \; x_r \right)
\qquad,
\label{vmema}
\end{equation}
where $v_0$ is a normalization constant. The self-consistent equations
fulfilled by the $K\times K$ matrix $V$ are immediately obtained and
reads
\begin{eqnarray}
V _{qr} &=& \frac 12 \delta _{qr}  \left( \hat g^{EMA} + \lambda +
i\epsilon -\frac qK -\frac 12 \right) + \frac 1{4K} -
\frac 1{4K^2} \sum _{s=1}^q \sum _{t=1}^r \left( U^{-1} \right) _{st}
\nonumber \\
U _{qr} &=&  V _{qr} +\frac 12 \delta _{qr}  \left( \hat g^{EMA} + \lambda +
i\epsilon -\frac {K+1-q}K -\frac 12 \right) + \frac 1{4K} 
\qquad ,
\label{selfv}
\end{eqnarray}
where $\delta$ denotes the Kronecker symbol. The saddle-point
equations of ${\cal F}^{EMA}$ with respect to $g^{EMA}$ and $\hat g^{EMA}$
respectively read
\begin{eqnarray}
\hat g^{EMA} &=&  \frac q{2(g^{EMA}-K)} \label{sphg} \\
g^{EMA} &=& \frac 1K\; \hbox{\rm Trace}\left[ \left( V + V^\# \right) ^{-1} 
\right] \qquad . \label{spg}
\end{eqnarray}
The saddle-point equation (\ref{rho1}) gives back
equation (\ref{spg}) while identity (\ref{sphg}) differs from (\ref{omega1})
since the EMA order parameters (\ref{ema},\ref{ema2}) are not exact
solutions of the extremization conditions over ${\cal F}$.

We have solved numerically the above self-consistency equations using
the following procedure. For given $K$, $q$ and $\lambda$, we start
with an arbitrary value of $g^{EMA}$, e.g. null, and compute $\hat
g^{EMA}$ from (\ref{sphg}). We then seek for matrices $V$ and $U$
fulfilling (\ref{selfv}). Among all possible solutions, we select the
matrix $V$ having all eigenvalues with positive imaginary parts so
that $v_M ^{EMA}$ (\ref{vmema}) is bounded for large arguments $\vec
x$.  The new value of $g^{EMA}$ is computed from (\ref{spg}). The
whole process is iterated until a fixed value of $g^{EMA}$ has been
found.  The density of states is then equal to the imaginary part of
$g^{EMA}$ divided by $-\pi$, see in order (\ref{ema}), (\ref{Grho})
and (\ref{spectre}).

\subsection{Spectrum, mobility edge and pseudo-gap}

The density of states obtained for $K=3$, $q=1$ and $q=5$ are shown
fig~2a and fig~3a respectively. The overall form of the EMA spectrum
coincides with the numerics. The quantitative agreement is of course
less precise at large disorder ($q=5$). We shall see in Section VI.C
how the theoretical spectrum may be improved using SDA.

On the right side of the spectrum, the mobility edge is accurately 
estimated~: $\lambda_+^{EMA} \simeq 2.13$
for $q=1$ and $\lambda_+^{EMA} \simeq 3.35 $ for $q=5$, compare to
Section III.A. Nevertheless, as expected from the presentation of EMA
in Section V.A, the oscillations of $p(\lambda )$ corresponding to 
non-extended states are not captured by the approximation. 
 
A similar situation takes place at small energy levels. 
EMA predicts a gap whose order of magnitude
($\lambda_-^{EMA} \simeq 0.064 $ for $q=1$ and $\lambda_-^{EMA} \simeq
0.46 $ for $q=5$) coincides with the findings of Section
III.A. However, from a qualitative point of view, EMA wrongly predicts
a vanishing density of state below $\lambda_-^{EMA}$.  This artifact
is well known in the case of the Bethe lattice \cite{abou,bray,thor} for which
EMA becomes exact.

We have analytically checked the scaling behaviour discussed in
Section  II.E in the simplest case $K=1$. In the
small $q , \lambda$ region the EMA resolvent is obtained from Section
V.B, 
\begin{equation}
g^{EMA} \simeq \frac{ -q + \sqrt{ q^2 - 8 \lambda}}{4 \lambda} 
\qquad . \label{poiu}
\end{equation}
The gap width is therefore given by $\lambda_-^{EMA} \simeq q^2/8$, in
agreement with the heuristic arguments of Section II.E indicating that
the size of the pseudo-gap scales as $O(q^2)$ for weak disorder.
Furthermore, the cross-over between ``one-dimensional'' behaviour
and rare events take place when the imaginary part of the resolvent
(\ref{poiu}) reaches its maximum, that is at $\lambda _{c.o.} \simeq 
q^2/4$. The corresponding density of states scales as $p_{c.o.} \simeq
1/q$, in quantitative agreement with the findings of Section II.E.
Notice also that the value of the ratio $\lambda _{c.o.}/\lambda _- \simeq 2$
is supported by the numerical results for $K=3$, see Section III.A.

\subsection{Relations of dispersion}

We have shown in Section IV.D that the autocorrelation function of
eigenvectors is accessible from the second largest eigenvalue $\Lambda
_2$ of $T$ (\ref{transf}).  Due to the similarity of $T^{EMA}$
(\ref{transfema}) with the evolution operator of a quantum oscillator,
we look for an eigenvector of the following form
\begin{equation}
v^{EMA}_2 [ x_1 , x_{2} , \ldots , x_{K} ] = v_1. \exp \left( \frac i2
\sum _{q,r=1}^K V _{qr} \; x_q \; x_r \right) \; \sum _{r=1}^K \gamma
_r x_r \qquad,
\label{v2ema}
\end{equation}
where $v_1$ is a normalization constant and $V_{qr}$ is the quadratic
form appearing in the maximal eigenvector $v_M$ and self-consistently
defined through (\ref{selfv}). An elementary calculation shows that
the above Ansatz indeed corresponds to an eigenvector of $T^{EMA}$ if
and only if the $K$ coefficients $\gamma _r$ arising in (\ref{v2ema})
fulfill the eigensystem
\begin{equation}
\tau \;\gamma _r = -\frac 1{2K} \sum _{q=r} ^K \sum _{s=1}^K
(U^{-1})_{qs} \; \gamma _s\qquad,
\label{v2suite}
\end{equation}
where $U$ has been defined in (\ref{selfv}) and $\tau =\Lambda _2 /
\Lambda _M$.  We are thus left with the diagonalization of a
$K\times K$ matrix whose complex eigenvalues $\tau _\ell$,
$\ell=1,\ldots ,K$ can be written as (\ref{rel3}),
\begin{equation}
\tau _\ell = \exp \left( -K \sigma _\ell + i 2 \pi K \phi _\ell \right
) \qquad . \label{taux}
\end{equation}
In doing so, we have at our disposal the first $K$ excited states of
$T$ and not only the first one. This is a peculiarity of the EMA
scheme, that was not expected from Section IV.D. The average resolvent
therefore obeys the asymptotic scaling (\ref{rel2})
\begin{equation}
\overline{G^{\cal S}_{jk} (\lambda + i\epsilon )} \sim \sum _{\ell=1}
^K C_\ell \; \exp (- \sigma _\ell \; d + i 2\pi \phi _\ell \; d )
\qquad .
\label{rel4}
\end{equation}
This large distance behaviour strengthens the data
modelling (\ref{lorfit}) used in Section III.C. Indeed, if relation
(\ref{rel4}) held for all $d$'s, the power spectrum of the
eigenvectors $|\tilde w (\theta ,\lambda )|^2$ would be well approximated by a
superposition of Lorentzian factors from equation (\ref{rel1}).

The wave numbers $\phi _\ell$ as well as the inverse autocorrelation
lengths $\sigma _\ell$ are plotted fig~6 and fig~7 vs. the eigenvalue
$\lambda$ for $K=3,q=1$ and $K=3,q=5$ respectively. The agreement
between the EMA prediction and the numerical results of Section III.B
is good. Notice that numerical points for the wave numbers $\phi _1$
and $\phi _2$ merge and lie between the theoretical branches when
$\lambda $ approaches $\lambda _+$. This phenomenon is simply due to
the coalescence of the first two peaks in (\ref{lorfit}) when $|\phi _2 -
\phi _1 | \simeq \sigma _1 / 2\pi$ arising at such energy
levels. 

In the extended part of the spectrum, EMA can therefore
be used to obtain some accurate effective relations of dispersion for
the eigenmodes of the Laplacian operator, even at strong disorder.

%
%

\section{Single defect approximation}

\subsection{Principle of the approximation}

The physical grounds of the single defect approximation have been
exposed in \cite{sda}. The basic idea is to treat exactly, via the
stationarity equations (\ref{rho1},\ref{omega1}) the interactions of a
single site, i.e. the defect with its surrounding neighbors belonging
to the effective medium \cite{thor} {\em taking into account the
fluctuations of the number of these neighbors} \cite{sda}.  This
approximation scheme amounts to perform one iteration of the
extremization condition on the free-energy functional ${\cal F}$ from
the EMA solution. Let us see how SDA works for the small-world
problem.

First, we consider the saddle-point equation for the effective potential
(\ref{omega1}). When inserting the EMA density (\ref{ema}) in the
r.h.s. of (\ref{omega1}), we obtain the SDA effective potential
\begin{equation}
\frac i2 \; \omega ^{SDA} (\vec x) = q \; \left[ \exp \left(
\frac {i \; \vec x ^{\; 2}}{2 (g^{EMA}- 2K)} \right) -1\right] \qquad ,
\label{sdaomega}
\end{equation}
up to irrelevant terms when $n \to 0$. This expression differs from 
the best quadratic potential obtained from EMA (\ref{ema2},\ref{sphg}).

To calculate the SDA density $\rho ^{SDA}$, the stationarity identity
(\ref{rho1}) requires the eigenvector of the transfer matrix
(\ref{transf}) in presence of the SDA potential
(\ref{sdaomega}). Such a calculation cannot be done
analytically. Remembering the interpretation of $\omega$
exposed in Section IV.D, we may however estimate the ratio of the 
probabilities that a particle be located 
in the $n$-dimensional abstract space at position $\vec x$ 
within each scheme of approximation by
\begin{eqnarray}
\frac{\rho ^{SDA} (\vec x)}{\rho ^{EMA} (\vec x) } &\simeq & \exp \left[
\frac i2 \; \omega ^{SDA}(\vec x) -\frac i2 \; \omega ^{EMA} (\vec x)  
\right] \nonumber \\
&\simeq & \sum _{c =0} ^\infty \frac{e^{-q}\; q^c}{c !} \;
\exp  \left[ \frac i2 \; \left( \frac{c}{g^{EMA} -2K} - \hat g^{EMA}
\right) \vec x ^{\; 2} \right]
\label{app}
\end{eqnarray} 
As in the simpler case of the random graph \cite{sda}, 
the anharmonicity of the effective potential $\omega ^{SDA}$ 
(\ref{sdaomega}) reflects the local 
fluctuations of connectivity. Using the above expression for $\rho
^{SDA}$, we can now study the localized states induced by the
geometrical defects we have observed in Section III.C. 

\subsection{Localized eigenvalues}

The second moment of the field $\vec x$ in the SDA
scheme reads, see (\ref{Grho},\ref{ema},\ref{app}),
\begin{eqnarray}
g^{SDA}  & =  & \lim_{n\rightarrow 0}
- \frac in \int d\vec x \; \rho ^{SDA}(\vec x ) \; \vec x ^{\; 2} \qquad ,
 \nonumber \\
&=  & \sum _{c =0} ^\infty \frac{e^{-q}\; q^c}{c !} \;
\left( \frac 1{g^{EMA}} +  \frac{c}{g^{EMA} -2K} - \hat g^{EMA}
 \right) ^{-1} \qquad .
\label{gsda}
\end{eqnarray} 
Above the mobility edge $\lambda_+ ^{EMA}$, the EMA density of states
vanishes. Thus, $g^{EMA} (\lambda )$ and  $g^{SDA}
(\lambda )$ are real numbers. However, the poles of  $g^{SDA}$ located
on the real axis can give rise to Dirac peaks in the density of states
\cite{sda}. Consider indeed such a singularity taking place at 
$\lambda _{sing} > \lambda_+ ^{EMA}$. Close to the singularity, 
we may write
\begin{equation}
g^{SDA} (\lambda ) \sim \frac{p_{sing}}{\lambda - \lambda _{sing} + 
i\epsilon } \qquad ,
\end{equation}
and obtain a contribution to the density of states equal
to $p_{sing} \; \delta ( \lambda - \lambda _{sing} )$ using
(\ref{spectre}). As seen from expression (\ref{gsda}), the poles of
$g^{SDA}$ form a discrete sequence of energy levels $\lambda ^{SDA} _c$
labelled by the integer number $c$. The values of $\lambda ^{SDA} _c$
and of the corresponding weights $p_c^{SDA}$ are listed Table~1 and
Table~2 for $K=3,q=1$ and $K=3,q=5$ respectively. The agreement with
the numerical values of Section III.A is good (see that Section for a
discussion of the reliability of numerical measures). Note that $c$
coincides with the coordination number of the centers of eigenmodes
measured in Section III.C. This strongly supports the statement that
defects are responsible for localization \cite{sda}. 

\subsection{Extended spectrum at strong disorder}

The EMA density of states for $K=3,q=5$ displayed fig~3a shows some
discrepancies with the numerics from the maximum density located at
$\lambda \simeq 1.5 - 2.0$ up to the mobility edge.  We have computed
the SDA density of states in the extended region from equations
(\ref{gsda}) and (\ref{spectre}). The result is shown fig~3a and
agrees in a much better way with numerics.

For completeness, we have also compute the SDA density of states at
low disorder $K=3,q=1$ in the extended region. The resulting spectrum
can hardly be distinguished from the EMA prediction, and is therefore
in very good agreement with numerics.

%
%

\section{Summary and Conclusion}

In this work, the spectral properties of the Laplacian operator on
mixed lattices, composed of a random graph structure superimposed to a
unidimensional ring have been investigated. Using some approximation 
schemes, we have been able to reproduce the numerical density of 
states with a good agreement and to obtain a quantitative description 
of the localized eigenstates. From this point of view, we have shown
that the single defect approximation introduced in the simpler context
of the random graph was also capable of giving reliable and precise
estimates of the localized energies in presence of a
finite-dimensional underlying geometry.

One of the most interesting aspects of the present work is the existence
of pseudo dispersion relations for disordered and thus non periodic
lattices. By studying the autocorrelation functions of eigenmodes, or
more precisely their power spectra we have shown that some features of
the pure lattice remain present even at strong disorder. It is
encouraging to notice that the effective medium approximation already
provides a very good description of the power spectra of the
eigenmodes. To what extent this result holds in higher dimensional
models is of course questionable.  

As emphasized in the introduction, the small-world architecture may be
of particular relevance to the field of polymers and more generally
chain-like systems with three-dimensional interactions,
e.g. proteins, DNA, ... With respect to the diffusion problem
studied here, realistic models of polymers or biological molecules  
have to take into account the internal degrees of freedom of the
monomers/constituents and their non-trivial interactions.
We hope nevertheless that the technical approach developed
in the present paper will be of use in these more complex cases. In
particular, the knowledge of the excited states of the self-consistent
transfer matrix should give access to the effective dispersion
relations of the eigenmodes. The latters are of primary importance to
understand the dynamical properties of such molecules \cite{rev,sim}, 
with possible applications to the relaxation of single DNA molecules 
\cite{har2}.

\vskip 1cm
{\bf Acknowledgments~:} I am particularly grateful to G. Biroli for numerous
comments and suggestions about this work. I also thank A.~Barrat, 
M.~Weigt and R.~Zecchina for useful discussions.

\end{document}